\documentclass[twocolumn,showpacs,preprintnumbers,amsmath,amssymb]{revtex4} 
 
\usepackage{graphicx}
\usepackage{dcolumn}
\usepackage{bm}
 
  \newcommand{\ZAMO}{\overline} 
  \newcommand{\EE}{\hat{E}}
  \newcommand{\LL}{\hat{L}}
  \newcommand{\ee}{\hat{e}} 
  \newcommand{\Lt}{\tilde{L}} 
  \newcommand{\Bp}{\mathcal{B}_{p}}   
  \newcommand{\Bf}{\mathcal{B}_{\phi}}

\begin{document} 
 
 
\title{ Constraints on the Evolution of Black Hole Spin due to 
        Magnetohydrodynamic Accretion}
\author{Masaaki Takahashi}%
\email{takahasi@phyas.aichi-edu.ac.jp} 
\affiliation{%
 Department of Physics and Astronomy, Aichi University of Education,  
 Kariya, Aichi 448-8542, Japan 
}
\author{Akira Tomimatsu}%
\email{atomi@gravity.phys.nagoya-u.ac.jp}
\affiliation{%
 Department of Physics, Nagoya University, Nagoya 464-8602, Japan  
}%
\date{\today}
 
\begin{abstract} 
 Stationary and axisymmetric ideal magnetohydrodynamic (MHD) accretion
 onto a black hole is studied analytically.  The accreting plasma
 ejected from a plasma source with low velocity must be super-fast
 magnetosonic before passing through the event horizon.  We work out and
 apply a trans-fast magnetosonic solution without the detailed analysis
 of the regularity conditions at the magnetosonic point, by introducing
 the bending angle $\beta$ of magnetic field line, which is the ratio of
 the toroidal and poloidal components of the magnetic field.  
 To accrete onto a black hole, the trans-magnetosonic solution has some 
 restrictions on $\beta$, which are related to the field-aligned
 parameters of the MHD flows.  One of the restrictions gives the
 boundary condition at the event horizon for the inclination of a
 magnetic field line.  We find that this inclination is related to the
 energy and angular momentum transport to the black hole.    
 Then, we discuss the spin-up/down process of a rotating black hole by 
 cold MHD inflows in a secular evolution timescale. 
 There are two asymptotic states for the spin evolution.  One is that
 the angular velocity of the black hole approaches to that of the 
 magnetic field line, and the other is that the spin-up effect by the
 positive angular momentum influx and the spin-down effect by the 
 energy influx (as the mass-energy influx) are canceled.  
 We also show that the MHD inflows prevents the evolution to the
 maximally rotating black hole.   
\end{abstract} 
 

\maketitle

\section{Introduction} 

 As central engines of active galactic nuclei, some compact X-ray
 sources and gamma-ray bursts, black hole accretion systems are
 generally accepted.  Recently, it is also confidently expected that the
 magnetic field is considered to play a very important role in the black
 hole accretion systems \citep{BB98}.
 In addition, Meier \cite{Meier05} proposed that the inner part of a
 black hole accretion inflow may enter a magnetically-dominated phase.
 Although black holes cannot support a magnetic field by themselves,
 they can be immersed in the magnetic field generated by currents in
 black hole accretion flows.   
 Then, above an equatorial disk and between the disk and a black hole,
 we can expect a magnetically-dominated ``magnetosphere'' like our Sun
 or pulsars, where large-scale (global) well-ordered magnetic fields
 extend to distant regions, and some part of the fields extend inwardly
 and enter the black hole.  In this magnetosphere, more magnetically
 important phenomena are expected.

 The research of most pulsating X-ray sources as accreting neutron stars 
 is based on the qualitative features of their spectra and their
 spin-up/down rates.  To discuss these period changes in these sources,  
 Ghosh \& Lamb \cite{GL79} studied the accretion torque on a magnetic
 neutron star from a Keplerian disk by solving the hydromagnetic
 equations along the dipole magnetic field lines.  
 In the black hole case, unfortunately, the spin of a black hole is
 unclear for every black hole candidate.   
 This is because many uncertain factors about the black hole--accretion 
 disk systems still remain to be understood; e.g, the black hole mass
 (the size of the black hole), the distribution of magnetic field and
 accreting plasma, and the radiative process by the accreting plasma,
 etc.  The spin of a black hole would be of basic importance for
 considering the environment around the black hole and the activities of
 the black hole--accretion disk systems.  Then, the understanding of the
 process of the hole's spin change would provide an important tool for
 exploring the astrophysical activity of a black hole accretion system.

 The evolution of black hole spin by accreting matter was first studied
 by Bardeen \cite{Bardeen70}. Assuming that the gas trickles down the
 event horizon from the inner-edge of an equatorial accretion disk,
 which was set to be located at the last stable circular orbit of the
 Kerr geometry, it was shown that a non-rotating black hole would always
 be span up during accretion. However, Thorne \cite{Thorne74} showed
 that photon capture by a gas accreting black hole would become
 important in the last stages of spin-up, where photons are emitted by
 the accretion disk, and reported a limiting value of $a/m \sim 0.9982$,  
 where $m$ and $a$ denote the mass and angular momentum per unit
 mass of the black hole, respectively.   
 This is because the hole's capture cross-section is greater for photons
 of negative angular momentum than photons of positive angular momentum.
 Then, the captured photons prevent complete spin-up to $a/m=1$.  
 When we consider a black hole immersed in a magnetosphere, the energy
 and angular momentum extraction due to a magnetic torque on the hole can
 be expected \citep{BZ77,TNTT90}.  The magnetic field also prevents the
 spin-up to $a/m=1$.

  The MHD inflow carries the total energy $E$ and angular momentum $L$
 into a black hole.  Then, the black hole will obtain mass $\delta m=E$
 and angular momentum $\delta J=L$, where $J=am$. 
 Although in this paper we assume that the background metric does not
 change by the accretion process (at least within our observational
 timescale), the black hole will evolve in a secular timescale. 
 That is, for example, the angular velocity, the spin and the surface
 area will be modified.  From the hole's angular velocity
 $\omega_{H}\equiv a/2mr_{H}$, where $r_{H}=m+\sqrt{m^2-a^2}$ is the
 horizon's radius, we can obtain the relation    
 \begin{equation} 
     \delta\omega_{H} = \frac{1}{2mr_{H}\sqrt{m^2-a^2}}
     \left[ L - \frac{a}{m}(m+r_{H}) E \right] \ . \label{eq:spin-BH}
 \end{equation}
 We may expect that the magnetic field lines connected to the lower
 latitude region of the event horizon contribute to the hole's spin-up
 and the field lines connected to the higher latitude (polar) region of
 the event horizon contribute the spin-down (e.g., see \citep{PV88}).
 To discuss the spin evolution, we need to construct a reasonable model
 of MHD accretion in a black hole magnetosphere.

 In this paper, we consider a stationary and axisymmetric black hole
 magnetosphere with a magnetized accretion disk, where plasma flows
 stream along magnetic field lines under the ideal MHD approximation.
 The global magnetic fields can be generated by current in the
 equatorial magnetized accretion disk and its off-equatorial plasma in
 the disk's corona region.  In this black hole magnetosphere, the
 magnetic field lines originated from the inner part of the accretion
 disk (within several gravitational radius) connect to the event
 horizon, and the fields from the outer part of the accretion disk
 extend to distant regions.  The plasma ejected from the disk surface
 must stream inwardly toward the event horizon along the black
 hole--disk magnetic field lines or  outwardly toward distant regions
 along the disk's open magnetic field lines; so that, we need both MHD
 ``inflow'' and ``outflow'' solutions, which are obtained from the
 general relativistic Bernoulli equation.

 The general relativistic Bernoulli equation for magnetized flows, which
 is the relativistic extension from the work by Weber \& Davis
 \cite{WD67}, was formulated by Camenzind \cite{Camenzind86} and
 Takahashi et al.\cite{TNTT90}. 
 We consider physically acceptable MHD accretion solutions that start 
 from a plasma source with low velocity, which is sub-Alfv\'{e}nic 
 velocity (or sub-slow magnetosonic when we consider a hot flow).    
 The physically acceptable accelerated inflow falls into the black hole 
 after passing through the Alfv\'{e}n point and the fast magnetosonic 
 point.  At the magnetosonic points, which are X-type (saddle type)
 critical points, a smooth transition from sub-magnetosonic to
 super-magnetosonic flows must occur.  Takahashi \cite{Takahashi02}
 investigated the critical conditions on the trans-magnetosonic flows in
 the Kerr geometry, and presented the restriction on the field-aligned 
 flow parameters for the appearance of the magnetosonic point on the 
 accretion solutions. In general, the task for finding a
 trans-magnetosonic solution is troublesome because the relations
 between the flow parameters are very complicated.  Therefore, in this
 paper, we will study the trans-fast magnetosonic MHD accretion solution
 without the detailed analysis of the regularity conditions at the
 magnetosonic point, by introducing the bending angle $\beta$ of
 magnetic field, which is defined as the ratio of the toroidal and
 poloidal components of the magnetic field.   
 The singular term in the equation of motion of MHD plasma is
 incorporated in this function.  We only assume that the function
 $\beta$ is regular at the magnetosonic point, and then we find that the
 regularity condition of the trans-magnetosonic flow solution is
 automatically satisfied.  Thus, we can easily find a trans-magnetosonic 
 solution, when the function $\beta(r,\theta)$ is given as a model.  
 In former treatments about trans-magnetosonic solution (including the 
 critical point analysis), the toroidal component of magnetic field is
 obtained by solving the relativistic Bernoulli equation under a 
 {\it given}\/ poloidal magnetic field.  Although, in this paper, 
 we assume a function $\beta$ and discuss the restrictions on it by
 field-aligned flow parameters (see Appendix~\ref{sec:beta}), we can say
 that our approach is basically equivalent to the discussion about the
 toroidal magnetic field along the trans-magnetosonic flow in the former
 treatments.

 Although we propose a new approach to solve the trans-magnetosonic
 solutions without the complicated regularity conditions at the
 magnetosonic points, the task for finding the function
 $\beta=\beta(r,\theta)$ is another difficult problem, which is not
 discussed here.  In this paper, a flux tube is considered, but the
 cross-section of the magnetic flux tube is not specified beforehand.  
 Of course, we should give a consistent function $\beta(r,\theta)$ along
 a magnetic field line for a black hole magnetosphere considered.    
 To do this, we need to treat the force-balance between magnetic field
 lines streaming the trans-magnetosonic flows.  However, by considering
 the relativistic Bernoulli equation, we can discuss the restrictions on
 the trans-magnetosonic flows in a black hole magnetosphere without a
 complete field configuration satisfying the force-balance equation 
 (the relativistic Grad-Shafranov (GS) equation \cite{Nitta-TT91}).  
 The GS equation would give another restrictions on flow solutions in 
 addition to the restrictions discussed in this paper.  
 This method for black hole accretion is the general relativistic
 extension from the special relativistic outgoing trans-magnetosonic
 flows by Tomimatsu \& Takahashi \cite{TT03}.

 In \S~\ref{sec:basic} we review the basic equations for
 trans-magnetosonic flows in Kerr geometry (see also
 \cite{TNTT90,Camenzind86}) and introduce the function $\beta$.  
 By assuming a smooth function of $\beta$, we show trans-magnetosonic
 flows without the regularity condition analysis.   
 In \S~\ref{sec:trans-f} we solve cold trans-fast magnetosonic flow
 solutions.  To fall into the black hole, the trans-fast magnetosonic
 solution must satisfy some conditions for the field-aligned flow
 parameters.  We show these restrictions (the necessary condition) in 
 \S~\ref{sec:restriction}.  
 Next, in \S~\ref{sec:bc-EH}, we present the boundary conditions at the
 event horizon for ingoing MHD flows.  We show that the magnetic field
 configuration restricts the signature of energy and angular momentum
 fluxes across the event horizon.  To specify the energy and angular
 momentum of MHD inflows, a model of the plasma source is necessary.
 But we can discuss the spin-up/down (as a secular evolution) of a
 rotating black hole by considering the restriction on MHD inflows at
 the event horizon.   
 Finally, we give brief remarks in \S~\ref{sec:conc}.

\section{ Basic Equations for MHD Flows }  \label{sec:basic}

 We consider stationary and axisymmetric ideal MHD flows in Kerr
 geometry. The background metric is written by the Boyer-Lindquist 
 coordinates with $c=G=1$, 
 \begin{eqnarray}
  ds^2 &=& \left( 1-\frac{2mr}{\Sigma} \right) dt^2 
   + \frac{4amr\sin^2\theta}{\Sigma} dt d\phi       \nonumber  \\
  & &  - \frac{{\cal A}\sin^2\theta}{\Sigma} d\phi^2
   - \frac{\Sigma}{\Delta} dr^2 - \Sigma d\theta^2  \ ,  
 \end{eqnarray}
 where $\Delta \equiv r^2 -2mr +a^2 $, 
 $\Sigma \equiv r^2 +a^2\cos^2\theta$,  
 ${\cal A} \equiv (r^2+a^2)^2 - \Delta a^2 \sin^2\theta$.  
 The particle number conservation is $(nu^\alpha)_{;\alpha}=0$,	where
 $n$ is the number density of the plasma and $u^\alpha$ is the fluid
 4-velocity.  The ideal MHD condition is $u^\beta F_{\alpha\beta}=0$,
 where $F_{\mu\nu}$ is the electromagnetic tensor.  The relativistic
 Polytropic relation is $P=K\rho_0^\Gamma$, where $\rho_0=nm_{\rm part}$
 is the rest mass density, $m_{\rm part}$ is the mass of the plasma
 particle and $\Gamma$ is the adiabatic index.  The equation of motion
 is ${T^{\alpha\beta}}_{;\beta}=0$. The energy-momentum tensor is given
 by $T^{\alpha\beta}= n\mu u^\alpha u^\beta - Pg^{\alpha\beta} 
     +(1/4\pi)[ {F^\alpha}_{\lambda}F^{\lambda\beta}
     +(1/4)g^{\alpha\beta}F^2 ]$, 
 where $\mu \equiv (\rho+P)/n$    
 is the relativistic enthalpy and $F^2\equiv F^{\mu\nu}F_{\mu\nu}$.  
 The magnetic and electric fields seen by a distant observer are defined by 
 $B_\alpha \equiv (1/2)\eta_{\alpha\beta\gamma\delta}k^\beta
 F^{\gamma\delta}$ and $E_\alpha \equiv F_{\alpha\beta}k^\beta$, 
 where $k^\alpha=(1,0,0,0)$ is the timelike Killing vector and 
 $\eta_{\alpha\beta\gamma\delta}\equiv
  (-g)^{1/2} \epsilon_{\alpha\beta\gamma\delta}$.  
 The poloidal component $B_p$ of the magnetic field seen by a lab-frame
 observer is defined by    
 \begin{eqnarray}
   B_p^2 &\equiv& - B^A B_A / G_t^{2} ~~~~(A=r,\theta)   \\ 
         &=&      - \left[ g^{rr}(\partial_r \Psi)^2  
         + g^{\theta\theta}(\partial_\theta \Psi)^2 \right]/{\rho_w^2} 
                                                      \ ,  
 \end{eqnarray}
 where $\Psi(r,\theta)$ is the magnetic stream function (the $\phi$
 component of the vector potential, $A_\phi$). 
 The poloidal component $u_p$ of the velocity is defined by 
 $u_p^2 \equiv -u^A u_A$.   
 [ Here, we set $u_p > 0$ for both ingoing flows ($u^r < 0$) and
 outgoing flows ($u^r>0$). ]

 The ideal MHD flows stream along the magnetic field lines
 (i.e., $\Psi(r,\theta)=$ constant lines), and have five field-aligned
 flow parameters; that is, the angular velocity of field lines
 $\Omega_F(\Psi)=-F_{tA}/F_{\phi A}$,   
 the number flux per magnetic flux $\eta(\Psi) = n u_p/B_p$, 
 the total energy $E(\Psi) = \mu u_t - \Omega_FB_\phi/(4\pi\eta) $, 
 the total angular momentum $L(\Psi)= -\mu u_\phi -B_\phi/(4\pi\eta) $
 and the entropy, which is related to $K(\Psi)$. 
 The relativistic Alfv\'{e}n Mach number $M$ is defined by 
\begin{equation}
      M^2\equiv \frac{4\pi\mu n u_p^2}{B_p^2} 
              = \frac{\hat{\mu} u_p }{\Bp} \ ,   \label{eq:mach_def}
\end{equation}
 where the term 
 $\Bp \equiv B_p/(4\pi\mu_{c}\eta) $  
 is introduced to nondimensionalize. 
 The relativistic enthalpy can be expressed in terms of $u_p$ and $B_p$
\begin{equation}
 \hat{\mu} \equiv \frac{\mu}{\mu_{c}} 
    = 1 + \left( \frac{\mu_{\rm inj}}{\mu_{c}} \right) \left(
    \frac{B_p}{u_p} \right)^{\Gamma-1}  
    = 1 + \mu_{\rm hot}\left( \frac{\Bp}{u_p} \right)^{\Gamma-1}   \ , 
\end{equation}
 where $\mu_{c}=m_{\rm part}$ and 
 $\mu_{\rm hot}\equiv (\mu_{\rm inj}/\mu_{c})(4\pi\mu_{c}\eta)^{\Gamma-1}$. 
 The term $\mu_{\rm inj}$ is evaluated at the plasma injection point by 
\begin{equation}
 \mu_{\rm inj} \equiv \frac{\Gamma K}{\Gamma-1}(\mu_{c}\eta)^{\Gamma-1}   
   =  \frac{\Gamma}{\Gamma-1} \frac{P_{\rm inj}}{n_{\rm inj}m_{\rm p}}  
      \left(\frac{u_p^{\rm inj}}{B_p^{\rm inj}} \right)^{\Gamma-1} \ . 
\end{equation}
 The toroidal magnetic field $B_\phi = (\Delta\sin\theta/\Sigma)
 F_{\theta r}$  can be expressed in terms of flow's parameters and the
 Alfv\'{e}n Mach number, and the non-dimensional toroidal magnetic field
 $\Bf$ is defined as  
\begin{equation}
  \Bf \equiv  \left( \frac{1}{\rho_w} \right) 
              \frac{B_\phi}{4\pi\mu_{c}\eta} 
           =  \frac{ G_\phi\EE  + G_t\LL }{\rho_w(M^2-\alpha)}  \ , 
  \label{eq:Bf}
\end{equation}
 where $\EE \equiv E/\mu_{c}$, $\LL \equiv L/\mu_{c}$, 
 $\alpha \equiv g_{tt}+2g_{t\phi}\Omega_F+g_{\phi\phi}\Omega_F^2$,
 $G_\phi \equiv g_{t\phi}+g_{\phi\phi}\Omega_F$,   
 and $G_t    \equiv g_{tt}+g_{t\phi}\Omega_F$.
 The locations of $\alpha(r;\Psi)=0$ give the inner and outer light
 surfaces for magnetic field lines of $\Omega_F=\Omega_F(\Psi)$. 
 When $M^2=\alpha$ at some location, it seems that the toroidal
 magnetic field diverges.  
 Such location is called the Alfv\'{e}n point, because the poloidal
 velocity of the flow equals the Alfv\'{e}n wave speed there.  
 To realize the physical trans-Alfv\'{e}nic flow, we also require the
 condition $L/E = -(G_\phi/G_t)_{\rm A}$ there (see \cite{TNTT90}),
 where the label ``A'' indicates quantities at the Alfv\'{e}n  radius.
 When we set $B_p>0$ (i.e., $\partial_\theta \Psi>0$) in the Northern  
 hemisphere and $B_p<0$ in the Southern hemisphere,  
 we obtain $\eta>0$ and $\eta<0$ in the respective hemispheres. Then, 
 the direction of the toroidal magnetic field $B_\phi$ is also reversed 
 for the equatorial plane.

 The poloidal equation (the relativistic Bernoulli equation) that gives
 the Mach number of the streaming plasma in the magnetosphere can be
 written as  
 \begin{equation}
     \ee^2 - \hat{\mu}^2 \alpha - M^4 (\alpha \Bp^2 + \Bf^2) =0 \ , 
                                                    \label{eq:pol_eq}
 \end{equation}
 where   $\ee\equiv \EE -\Omega_F \LL$.   
 The differential form of the poloidal equation~(\ref{eq:pol_eq}) can be
 written as  
\begin{equation}
  (\ln u_p)' = \left[ \ln \left( \frac{\Bp}{\hat{\mu}} \right) \right]' 
              - \frac{M^4}{ H } 
               \left[ \alpha \left( \frac{\Bp}{\hat{\mu}} \right)^2 
              + \left( \frac{\Bf}{\hat{\mu}} \right)^2 \right]' 
              - \frac{1}{ H } \alpha'  \ .     \label{eq:diff_up}
\end{equation}
 where 
 $ { H } \equiv   (2/\hat{\mu}^2) \left(
        \hat{e}^2 - \hat{\mu}^2\alpha - C_{\rm sw}^2\hat{e}^2 \right) $, 
 $ C_{\rm sw} \equiv a_{\rm sw} / (1-a_{\rm sw}^2)^{1/2} $ is the sound 
 four-velocity and $a_{\rm sw} = [(\Gamma-1)(1-\hat{\mu}^{-1})]^{1/2} $ 
 is the sound three-velocity.  The prime is a derivative along a stream
 line $\partial_r + (B^\theta/B^r)\partial_\theta$ .  
 It seems that the singularities at the Alfv\'{e}n point and the fast
 and slow magnetosonic points are removed from the differential form of
 the poloidal equation, but we should note that the critical behavior
 is included in the term of $(\Bf^2)'$. In fact, by substituting $\Bf$
 expressed as equation (\ref{eq:Bf}) for equation~(\ref{eq:diff_up}), 
 we obtain the traditional expression; that is, 
 $(\ln u_p)' = {\cal N}/{\cal D}$, where the numerator ${\cal N}$ and
 the denominator  
 ${\cal D} \propto (u_p-u_{\rm AW})^2(u_p-u_{\rm FM})(u_p-u_{\rm SM})$
 are the functions of $M^2$, $r$ and $\Psi$ with field-aligned flow
 parameters  (see \cite{TNTT90} and \cite{Takahashi02} for the
 analysis at the magnetosonic points). 
 The terms $u_{\rm AW}$, $u_{\rm FM}$, $u_{\rm SM}$ are the Alfv\'{e}n
 wave speed, the fast and slow magnetosonic wave speeds, respectively.

 When we assume a magnetic flux function $\Psi=\Psi(r,\theta)$ and try
 to solve the poloidal equation, we need to specify a set of five 
 field-aligned parameters satisfying the critical conditions at the
 X-type magnetosonic points and the Alfv\'{e}n point. To obtain a
 physical trans-magnetosonic flow solution, the fine-tuning of the
 parameters is required. In general this task is very complicated. 
 However, we now propose a new analytical method to study the
 trans-magnetosonic flows without the critical conditions.  We can
 relate the toroidal magnetic field to the poloidal magnetic field by
 defining the bending angle of a magnetic field line as  
 \begin{equation}
    \beta(r,\theta)\equiv \frac{\Bf}{\Bp} \ .     \label{eq:beta_def}
 \end{equation}
 Then, in the differential form of the poloidal
 equation~(\ref{eq:diff_up}), $(\beta^2)'$ takes the place of
 $(\Bf^2)'$.  When $\beta$ is a smooth function at the magnetosonic
 points, which is a natural situation on accretion problems, 
 there is no need to analyze the regularity condition there.  
 That is, in the $r$-$u_p$ plane, the inclination $(u_p)'$ of a flow 
 solution is determined with a finite value anywhere.  
 At the event horizon, the toroidal magnetic field becomes 
 $ B_{\phi}^{H} = (-g_{\phi\phi}/\Sigma)_{H}^{1/2} (\omega_{H}-\Omega_F)
   (\partial_\theta\Psi)_{H} $  and $\Psi(r_{H}, \theta)=$ finite, which
 are the boundary conditions there. The label ``H'' indicates quantities
 at the event horizon.  Then, we find the condition 
 $\beta_{H} = (-g_{\phi\phi})_{H}^{1/2} (\omega_{H}-\Omega_F) $.  
 Thus, we can obtain physical accretion solutions passing through the 
 event horizon with a finite Mach number.

 We also introduce the poloidal electric--to--toroidal magnetic field
 amplitude ratio $\xi^2$ seen by a zero angular momentum observer
 (ZAMO) as (see also \cite{TT03}) 
\begin{equation}
   \xi^2(r,\theta) \equiv \left( \frac{\ZAMO{E}_p}{\ZAMO{B}_T} \right)^2
   \ = \ \frac{G_\phi^2}{\rho_w^2}
         \left( \frac{\ZAMO{B}_p}{\ZAMO{B}_T} \right)^2 \ ,
                                                       \label{eq:xi-ZAMO} 
\end{equation}
 where we use the following relations 
 $ \ZAMO{E}_p^2 \equiv -\ZAMO{E}^A\ZAMO{E}_A 
   = (G_\phi/\rho_w)^2\ZAMO{B}_p^2 $, 
 $ \ZAMO{B}_p^2 \equiv -\ZAMO{B}^A\ZAMO{B}_A = \alpha_Z^2 B_p^2 $, 
 $ \ZAMO{B}_T^2 \equiv -\ZAMO{B}^\phi\ZAMO{B}_\phi = (B_\phi/\rho_w)^2 $.   
 The magnetic and electric fields seen by a ZAMO are defined by 
 $\ZAMO{B}_\alpha \equiv (1/2)\eta_{\alpha\beta\gamma\delta} 
  h^\beta F^{\gamma\delta}$ and 
 $\ZAMO{E}_\alpha \equiv F_{\alpha\beta} h^\beta$, where
 $h^\beta=(h^t_{Z}, 0, 0, h^\phi_{Z})= \alpha_{Z}^{-1}(1, 0, 0, \omega)$
 is the four-velocity of a ZAMO seen by a distant observer
 (see \cite{BPT72}).   
 The term $\alpha_{Z}\equiv 1/(g^{tt})^{1/2} = (\Sigma\Delta/A)^{1/2}$   
 is the lapse function and $\omega=-g_{t\phi}/g_{\phi\phi}$ is the
 angular velocity of a ZAMO with respect to a distant observer.  
 Then, we obtain the relation 
 $\xi^2 = - g_{\phi\phi}(\Omega_F-\omega)^2/\beta^2$. 
 At the event horizon, we obtain $\xi^2_{H}=1$.  
 Although the definitions and formalisms introduced in this section 
 are available for hot MHD flows ($P\neq 0$), in the following
 section for the sake of simplicity let's consider the cold MHD flows
 ($P=0$).

\section{ Cold Trans-Fast Magnetosonic Flow Solutions }
\label{sec:trans-f}

 For a cold MHD flow ($\hat{\mu}=1$) the poloidal equation
 (\ref{eq:pol_eq}) is reduced to    
\begin{equation}
   A M^4 -2B M^2 + C =0 \ , 
\end{equation}
 where 
\begin{eqnarray}
  A &=& \left( z -\frac{1}{\beta^2} \right)
       \frac{1}{\rho_w^2}( G_\phi\EE +G_t\LL )^2  \ ,  \\
  B &=& \ee^2 - \alpha                            \ ,  \\
  C &=& \alpha( \ee^2 - \alpha )               
\end{eqnarray}
 with  $ z \equiv -( k + 1 )\rho_w^2/(G_\phi\EE +G_t\LL)^2$, 
 and $ k \equiv (g_{\phi\phi}\EE^2 + 2g_{t\phi}\EE\LL +
 g_{tt}\LL^2)/\rho_w^2$.   
 Then, the Alfv\'{e}n Mach number of the flow is solved by  
\begin{equation}
   M^2_{\pm}(r, \theta) = \frac{ B \pm ( B^2-AC )^{1/2} }{ A } \ ,  
\end{equation}
 where $M_{+}^2$ denotes super-Alfv\'enic solution ($M^2>\alpha$) and 
 $M_{-}^2$ denotes sub-Alfvenic solution ($M^2<\alpha$). 
 The discriminant of the quadratic equation is denoted by 
\begin{equation}
   B^2-AC = \frac{ \alpha + \beta^2 }{\rho_w^2 \beta^2 }  
            (G_\phi\EE +G_t\LL)^2 (\ee^2 - \alpha)  \ .   \label{eq:discri}
\end{equation}

 Although the locations of the Alfv\'{e}n point are given by the flow
 parameters $L/E$ and $\Omega_F$ for each magnetic field line $\Psi=$
 constant, the distribution of the Alfv\'{e}n surfaces in a black hole
 magnetosphere is obtained when the magnetic field distribution
 $\Psi=\Psi(r, \theta)$ and the boundary conditions at the plasma source
 $E=E(\Psi)$, $L=L(\Psi)$ and $\Omega_F=\Omega_F(\Psi)$ are
 specified. The plasma source (the plasma injection point) is located
 within the sub-Alfv\'{e}nic region, where the poloidal flow velocity is
 less than the Alfv\'{e}n wave speed, and the neighborhood of the event
 horizon and the distant region from the plasma source are in the
 super-Alfv\'{e}nic region, where the poloidal flow velocity is greater
 than the Alfv\'{e}n wave speed.  The accretion/wind solution is denoted
 by $M^2=M^2_{-}$ in the sub-Alfv\'{e}nic region and by $M^2=M^2_{+}$ in
 the super-Alfv\'{e}nic region. Both branches of the solutions always
 connect smoothly at the Alfv\'{e}n point, where $B^2-AC=0$; that is, 
 $(M^2_{+})_{\rm A}=(M^2_{-})_{\rm A}=\alpha_{\rm A}$. 
 Furthermore, at the light surfaces, we see that $(M^2_{-})_{L}=0$,
 while $(M^2_{+})_{L} = 2\ee^2/A_{L}$ is finite except the case that
 $A_{L}=0$ accidentally.  The label ``L'' indicates quantities at the
 light surfaces.

 If the coefficient $A$ becomes zero at some location for a given
 magnetic field line specified by $\xi^2=\xi^2(r; \Psi)$, the Mach
 number $M^2_{+}$ of the super-Alfv\'{e}nic flow solution diverges
 there, while $M^2_{-}$ has a finite value in the sub-Alfv\'{e}nic
 region.  This means that such a magnetic field line under a given
 flow's parameter set is {\it unphysical}\/ in the super-Alfv\'{e}nic
 region.   
 In fact, from equations (\ref{eq:mach_def}) and (\ref{eq:Bf})  both the
 toroidal magnetic field $\Bf$ and the poloidal magnetic field $\Bp$
 should vanish at the location of $A=0$, where $M^2=M^2_{+}\to\infty$.   
 (Note that, although $(\Bf)_{A=0} = (\Bp)_{A=0} = 0$, the ratio $\beta$
 must have a finite value of $\beta^2=1/z$ at $A=0$.)  Thus, such a
 solution (where $A=0$ in the super-Alfv\'{e}nic region) can not be
 accepted as a solution of trans-magnetosonic accretion/wind.      
 To obtain a physically acceptable MHD flow, the magnetic field
 configuration characterized by such a function $\xi=\xi(r,\theta)$
 should be modified to avoid the appearance of $A=0$ location, or
 another set of flow parameters should be selected at the plasma source.

 For the inflow (or outflow) streaming toward the black hole (or distant 
 regions), if $A=0$ on the sub-Alfv\'{e}nic flow, the solution has a
 finite Mach number across this location.  However, if $A=0$ in the
 super-Alfv\'{e}nic flow region, the Mach number diverges there; 
 the super-Alfv\'{e}nic flow is only available in the $A>0$ region.    
 Thus, $A>0$ in the super-Alfv\'{e}nic region is an {\it necessary
 condition}\/ for trans-magnetosonic accretion/wind solutions.  
 On the other hand, the $A<0$ region is the forbidden region on the
 super-Alfv\'{e}nic magnetosonic flow solution, where $M^2_{+}<0$ is
 obtained.

\begin{figure}[t] 
  \includegraphics[scale=0.45]{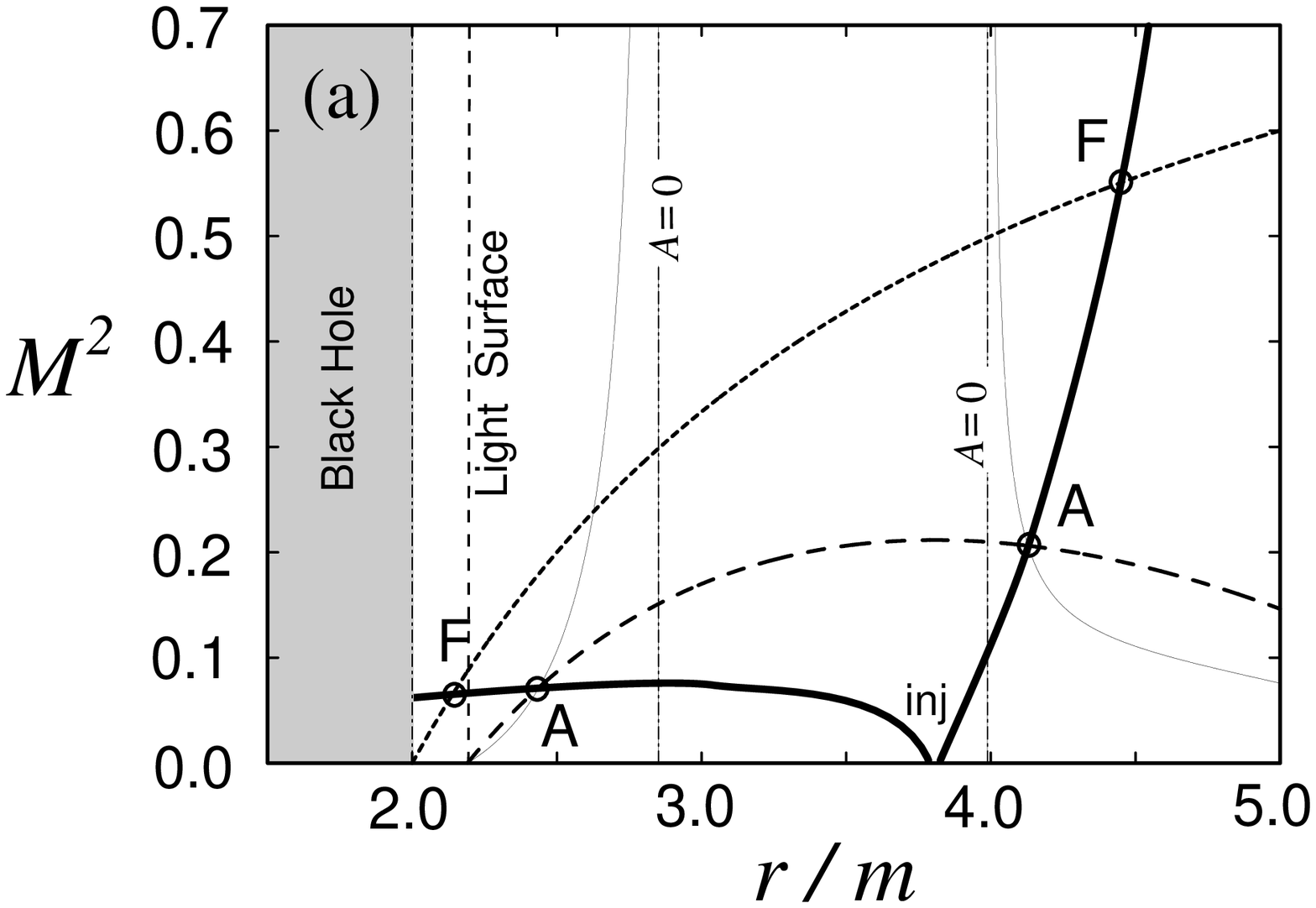}  \hspace{0.3cm}
  \includegraphics[scale=0.45]{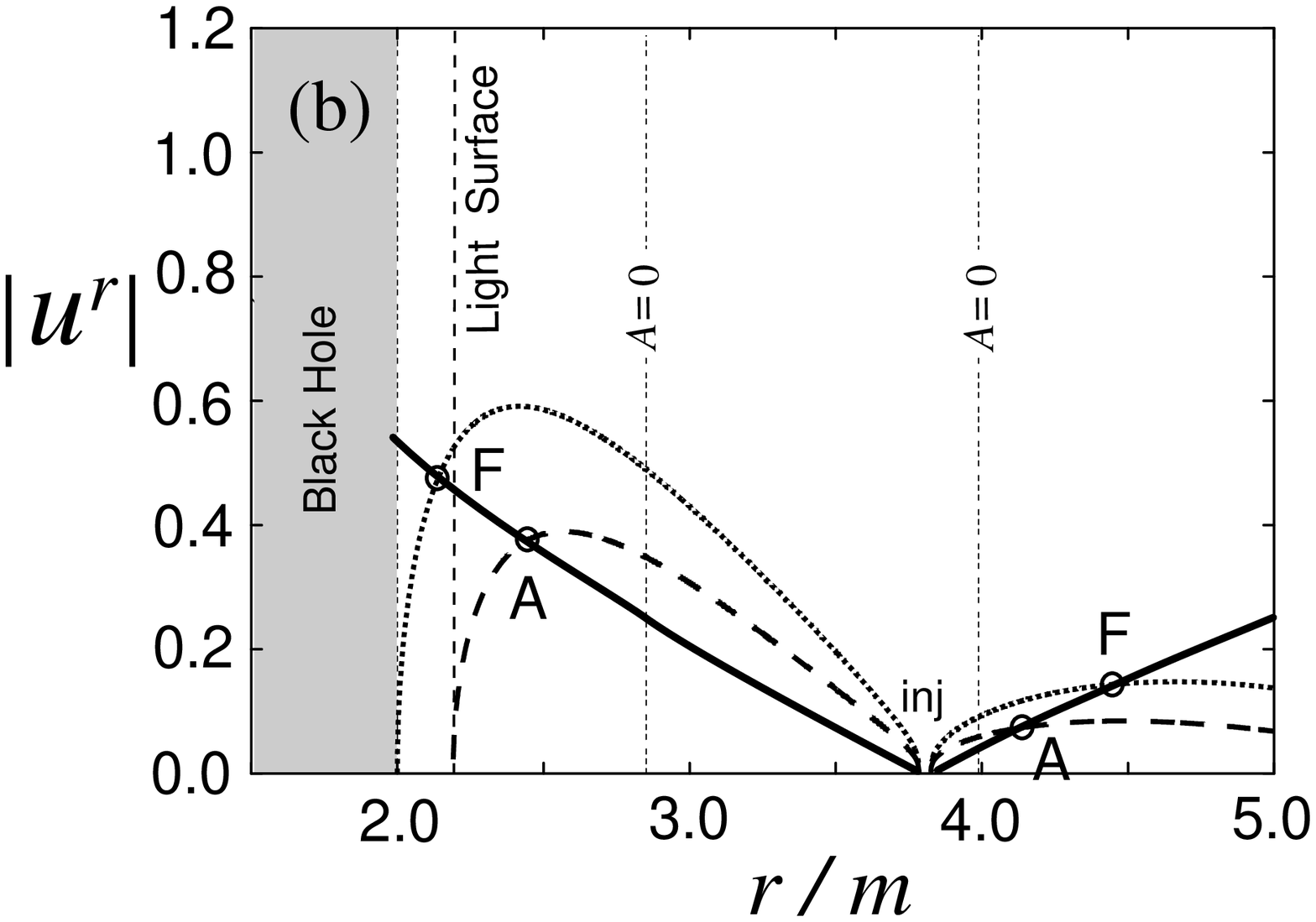}
 \caption{
 Transfast magnetosonic inflow and outflow solutions ({\it thick bold
 curves}) started from the separation surface with zero velocity
 (labeled ``inj'').     
 (a) The square of the  Alfv\'en Mach number $M^2$ vs. radius 
 $r/m$ and (b) the radial 4-velocity of the flow $u^r$ vs. radius $r/m$.
 The solution is plotted on the equatorial plane ($\theta=\pi/2$) in
 Schwarzschild geometry ($a=0$), where the magnetic field configuration 
 is assumed with $\xi(r,\theta)=1.0$. The flow parameters are given as
 $\EE=1.15$, $m\Omega_F= 0.7\Omega_{\rm max}$, $\LL\Omega_F=0.6$, 
 where $\Omega_{\rm max}$ is the maximum value of $\Omega_F$, and is
 given as the double roots of $\alpha=0$. 
 (If $\Omega_F > \Omega_{\rm max}$, no light surfaces exist along a
 magnetic field line considered.)   
 The broken curve shows the $M^2=M^2_{\rm AW}$ curve ({\it left}\/) and
 the $|u^r|=u^r_{\rm AW}$ curve ({\it right}\/), and the dotted curve
 shows the $M^2=M^2_{\rm FM}$ curve ({\it left}\/) and the
 $|u^r|=u^r_{\rm FM}$ curve ({\it right}\/).  The crossing points of
 these curves with the flow solution labeled by ``A'' and ``F'' are the
 Alfv\'en and fast magnetosonic points, respectively. The thin curves
 started from the light surface and approached to the $A=0$ line are
 unphysical solutions of the quadratic equation.     
 }
 \label{fig:acc}
\end{figure}

\begin{figure}[h] 
  \includegraphics[scale=0.45]{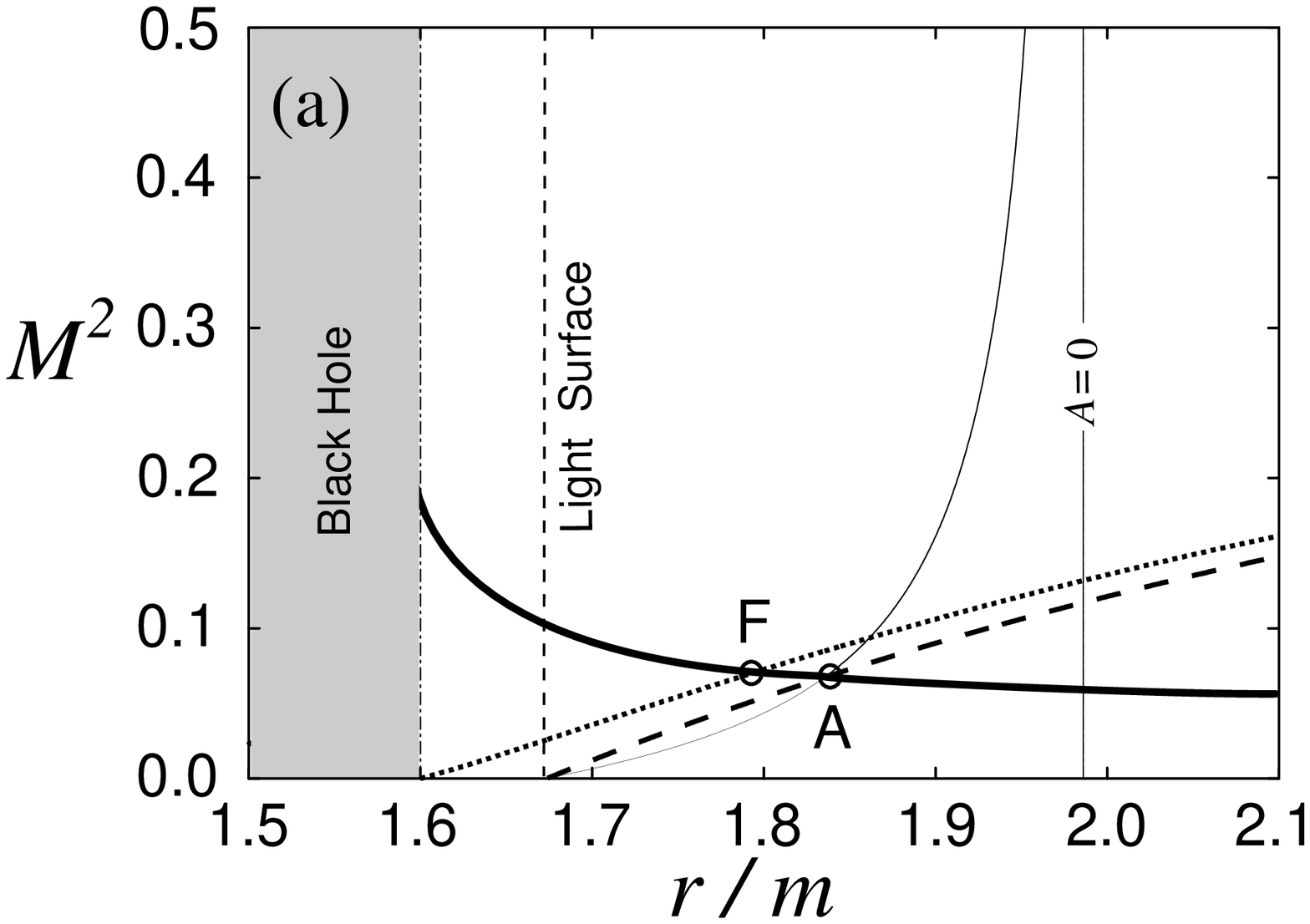} \hspace{0.5cm}
  \includegraphics[scale=0.45]{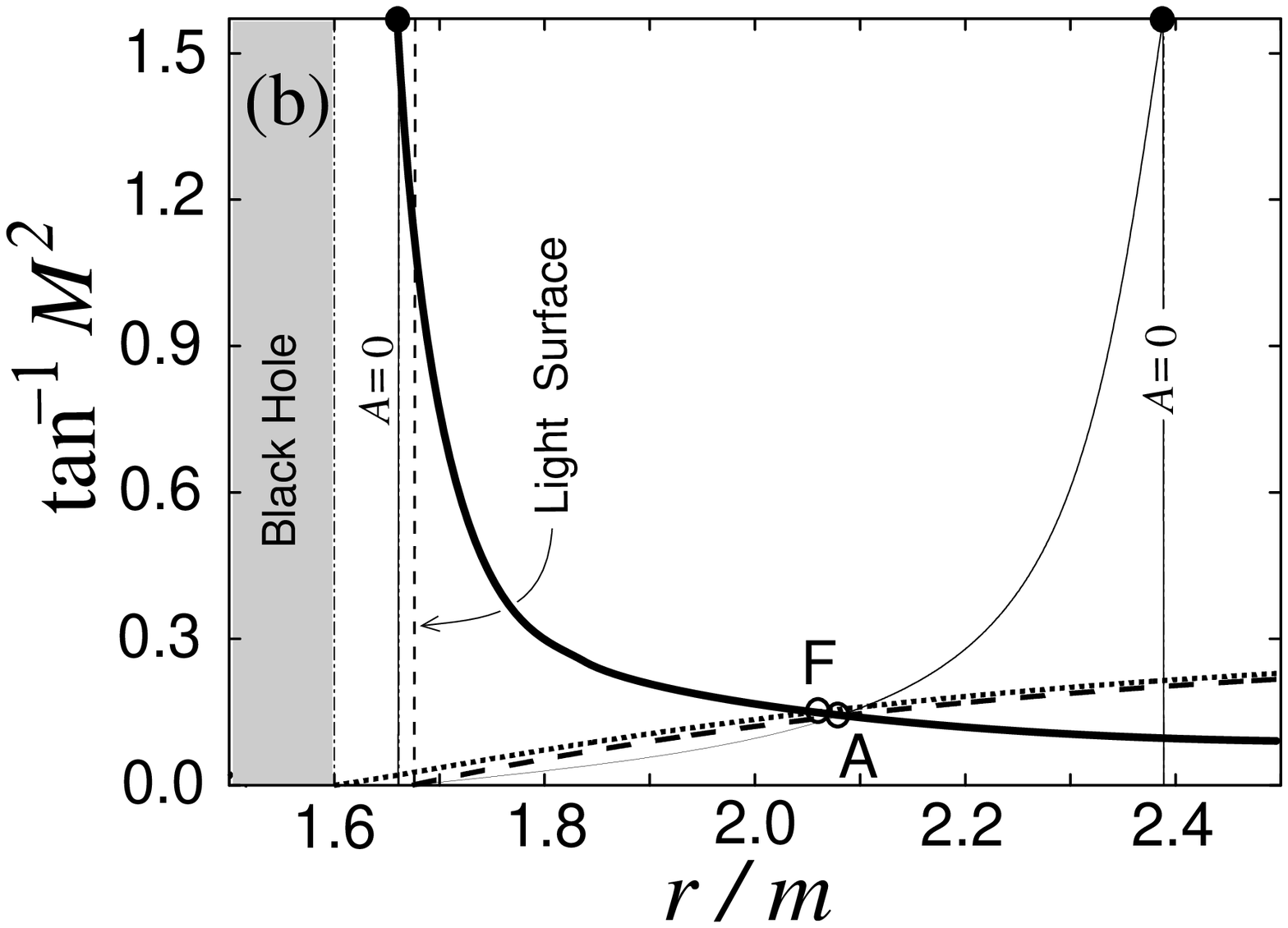}
 \caption{
   The square of the Alfv\'en Mach number $M^2$ vs. radius $r/m$ for 
   transfast magnetosonic inflow solutions ({\it thick bold curve}) on 
   the equatorial plane ($\theta=\pi/2$) in Kerr geometry ($a=0.8m$).
   The flow parameters are given as $\EE=1.0$, $m\Omega_F=0.5\Omega_{\rm
   max}$, $\LL\Omega_F=-0.5$ ({\it left}\/) and $\LL\Omega_F=0.09$ ({\it
   right}\/).  The magnetic field configuration is assumed by equation
   (\ref{eq:xi_IIin}) discussed in \S \ref{sec:beta}.  
   The right-hand figure shows an {\it unphysical solution}\/, where the 
   Alfv\'{e}n  Mach number $M^2$ diverges at the location of $A=0$. 
   The black spots ({\it right}\/) indicate $M^2=\infty$.  
   The dotted curve shows the $M^2=M^2_{\rm AW}$ curve, and the solid
   curve shows the $M^2=M^2_{\rm FW}$ curve. The crossing points of
   these curves with the flow solution labeled by ``A'' and ``F'' are
   the Alfv\'en and fast magnetosonic points, respectively.  The thin
   curves started from the light surface and approached to the $A=0$
   line are unphysical as inflow solutions.   
 }
 \label{fig:acc-Kerr}
\end{figure}

 For the cold MHD flow, we define two characteristic Alfv\'{e}n Mach
 numbers related to the Alfv\'{e}n and fast magnetosonic wave speeds
 (see \cite{TNTT90}) 
 \begin{eqnarray}
  M_{\rm AW}^2 (r,\theta)  &=&  \alpha  \ ,           \label{eq:maw}  \\
  M_{\rm FM}^2 (r,\theta)  &=& \alpha + \beta^2  \ .  \label{eq:mfw}  
 \end{eqnarray} 
 The locations of $M^2_\pm = M^2_{\rm AW}$ and $M^2_\pm = M^2_{\rm FM}$
 indicate the Alfv\'{e}n point and the fast magnetosonic point,
 respectively. Just on the event horizon, we obtain $M_{\rm FM}^2 = 0$. 
 Figure~\ref{fig:acc} shows the ingoing and outgoing flow solutions;
 both solutions are started from the separation surface, which separates
 the gravitational force dominated region and the centrifugal force
 dominated region (see~\cite{TNTT90}), with zero poloidal velocity.
 These solutions cross the $M^2 = M^2_{\rm AW}$ and $M^2=M^2_{\rm FM}$
 curves in this order (Fig.~\ref{fig:acc}a), and the 
 $|u^r| = u^r_{\rm AW} \equiv \sqrt{\Sigma/\Delta} (B_p^2/4\pi\mu n)
 \alpha$ and $|u^r| = u^r_{\rm FM} \equiv \sqrt{\Sigma/\Delta}
 (B_p^2/4\pi\mu n) (\alpha + \beta^2)$ curves (Fig.~\ref{fig:acc}b).
 Because $\eta =$ constant through the flow solution, the number density
 $n$ diverges at the injection point, where $u^r=0$, so that our
 definite radial Alfv\'en wave speed and the fast magnetosonic wave
 speed become zero, 
 $(u^r_{\rm AW})_{\rm inj} = (u^r_{\rm FM})_{\rm inj} = 0$, at the
 injection point.

 Note that the function $\xi(r,\theta)$ gives the distribution of the
 cross section of the magnetic flux tube in the poloidal plane. To plot 
 the flow solution as shown in Figure~\ref{fig:acc}, we need to specify
 the stream line of the flow; in Figure~\ref{fig:acc}, the flow
 streaming along the equatorial plane is plotted, but the magnetic flux 
 tube determined by the function $\xi$ is not a conical shape. 
 Figure~\ref{fig:acc-Kerr} shows the ingoing flow solutions for rapidly
 rotating Kerr black hole cases.  Each inflow enters the event horizon
 with a finite Mach number (Fig.~\ref{fig:acc-Kerr}a) or breaks at the
 $A=0$ location (Fig.~\ref{fig:acc-Kerr}b), where the Mach number of the
 trans-fast MHD inflow diverges. 
 In Figures~\ref{fig:acc} and~\ref{fig:acc-Kerr}, the functions 
 $\xi(r,\theta)$ are considered as simple models that satisfied
 conditions specified at some characteristic radii (the details of the
 conditions will be presented in Appendix~\ref{sec:beta}). The
 general properties of trans-magnetosonic flows discussed here do not
 depend on detail of the functional form of $\xi(r,\theta)$.

 The poloidal velocity of a cold MHD flow can be expressed in terms of
 $M^2$ and $\beta^2$ (or $\xi^2$) with the conserved quantities as    
\begin{equation}
    u_p^2 = \Bp^2 M^4 
          = \frac{\beta^2 ( G_\phi\EE +G_t\LL )^2 M^4}
                 {\rho_w^2 (M^2-\alpha)^2}  \ .  \label{eq:pol_v}
\end{equation}
 When we consider a solution for accretion onto a black hole, we apply 
 $M^2=M^2_{-}$ in the sub-Alfv\'enic region of $r>r_{\rm A}$ and
 $M^2=M^2_{+}$ in the super-Alfv\'enic region of $r<r_{\rm A}$ (see
 Fig.~\ref{fig:acc}b).  By using the poloidal equation
 (\ref{eq:pol_eq}), equation (\ref{eq:pol_v}) can be reduced to  
\begin{equation}
    u_p^2 = \frac{ \ee^2-\alpha }{\alpha + \beta^2} \ ,  \label{eq:pol_u}
\end{equation}
 which corresponds to a physical trans-Alfv\'enic accretion solution. 
 From the requirement of $u_p^2>0$, we find the minimum energy
 $\EE_0(r,\theta)$ at each place between the inner and outer light
 surfaces. The energy $E$ should be taken as $E>E_0(r;\Psi)$ along the
 flow (between the plasma source and the event horizon). 
 Note that, at the location of $A=0$ in the super-Alfv\'{e}nic region
 (if such a point exists), the poloidal velocity has a finite value,
 although the Alfv\'{e}n Mach number diverges.  However, the poloidal
 and toroidal components of the magnetic field vanishes there, while
 $\beta$ has a finite non-zero value, as mentioned in the first half of
 this section.

\section{ Restrictions on MHD Flows }           \label{sec:restriction}

\subsection{ Trans-Alfv\'{e}nic Flow }               \label{sec:alfven}

 Along a magnetic field line $\Psi(r,\theta) =$ constant, the location   
 of the Alfv\'{e}n point ($r_{\rm A}, \theta_{\rm A}$) is given by 
 $M^2 = \alpha$ and 
 \begin{equation}
         \Lt\Omega_F = Y_{\rm A}      \              \label{eq:alfven}
 \end{equation}
 with the definition of a function 
 \begin{equation}
         Y \equiv -G_\phi\Omega_F/G_t \ , 
 \end{equation}
 where $\Lt \equiv L/E $.  Although $\Lt\Omega_F$ is a function of
 $\Psi$, hereafter, we can regard $\Lt\Omega_F$ as a function of 
 $r_{\rm A}$ along the magnetic field line considered; note that
 $\theta_{\rm A} = \theta_{\rm A}(r_{\rm A}; \Psi)$.   
 In the black hole magnetosphere, two surfaces of the Alfv\'{e}n points  
 (i.e., the Alfv\'{e}n surfaces) for inflow and outflow are distributed
 between the inner and outer light surfaces given by the relation
 $\alpha = 0$.    
 For an ingoing MHD flow, the region between the Alfv\'{e}n surface and 
 the event horizon is the super-Alfv\'{e}nic region, while the region
 between the plasma source and the Alfv\'{e}n surface is the
 sub-Alfv\'{e}nic region.  

\begin{figure*}
  \includegraphics[scale=0.75]{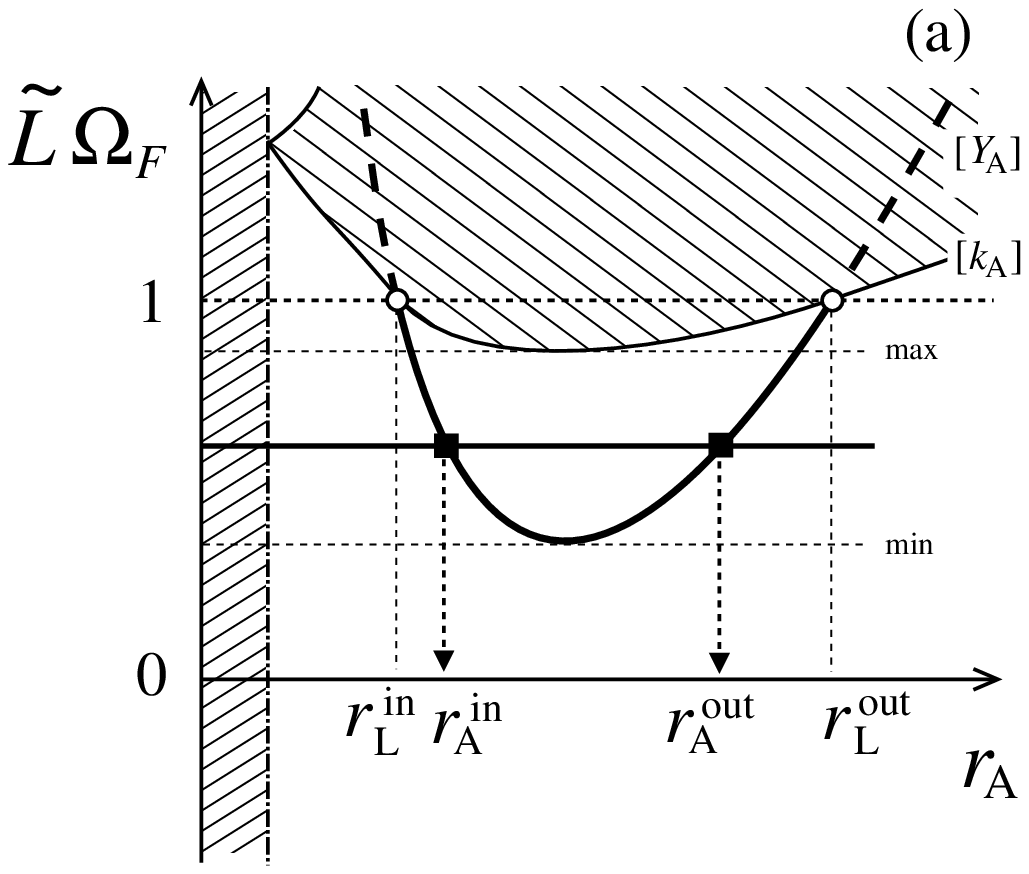} \hspace{1.0cm}
  \includegraphics[scale=0.75]{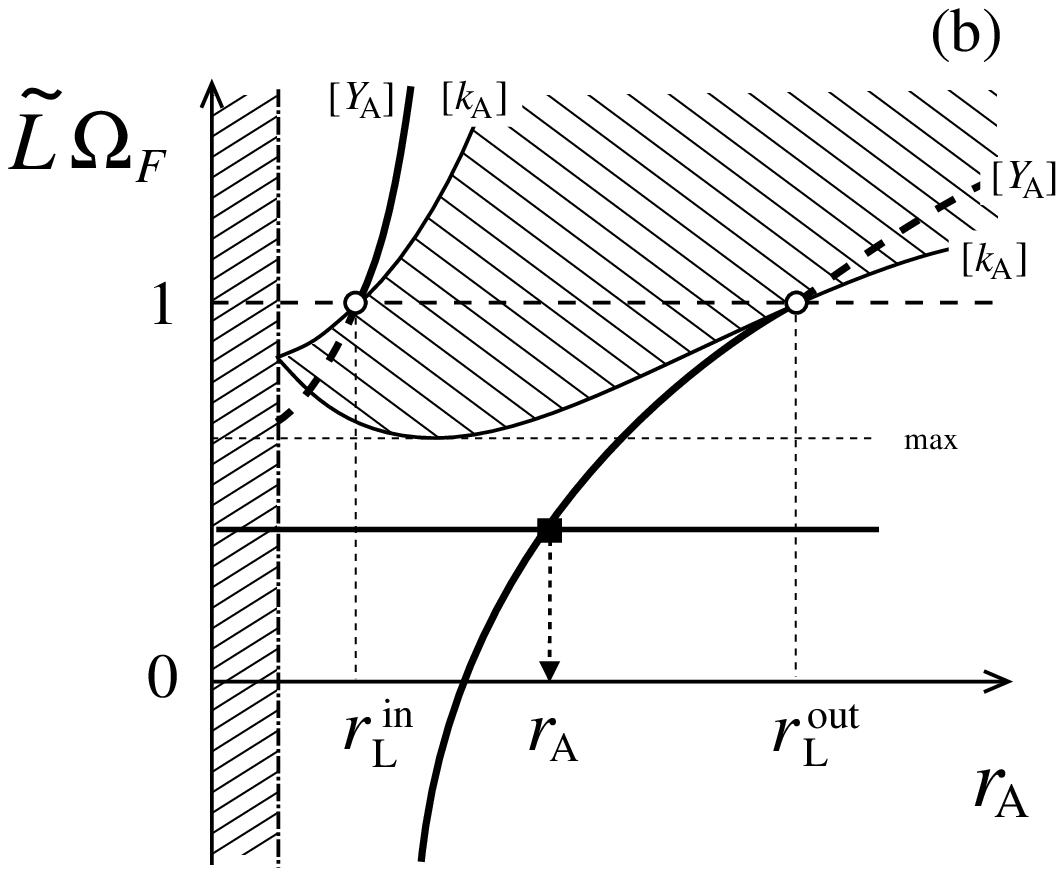} \\ 
  \includegraphics[scale=0.75]{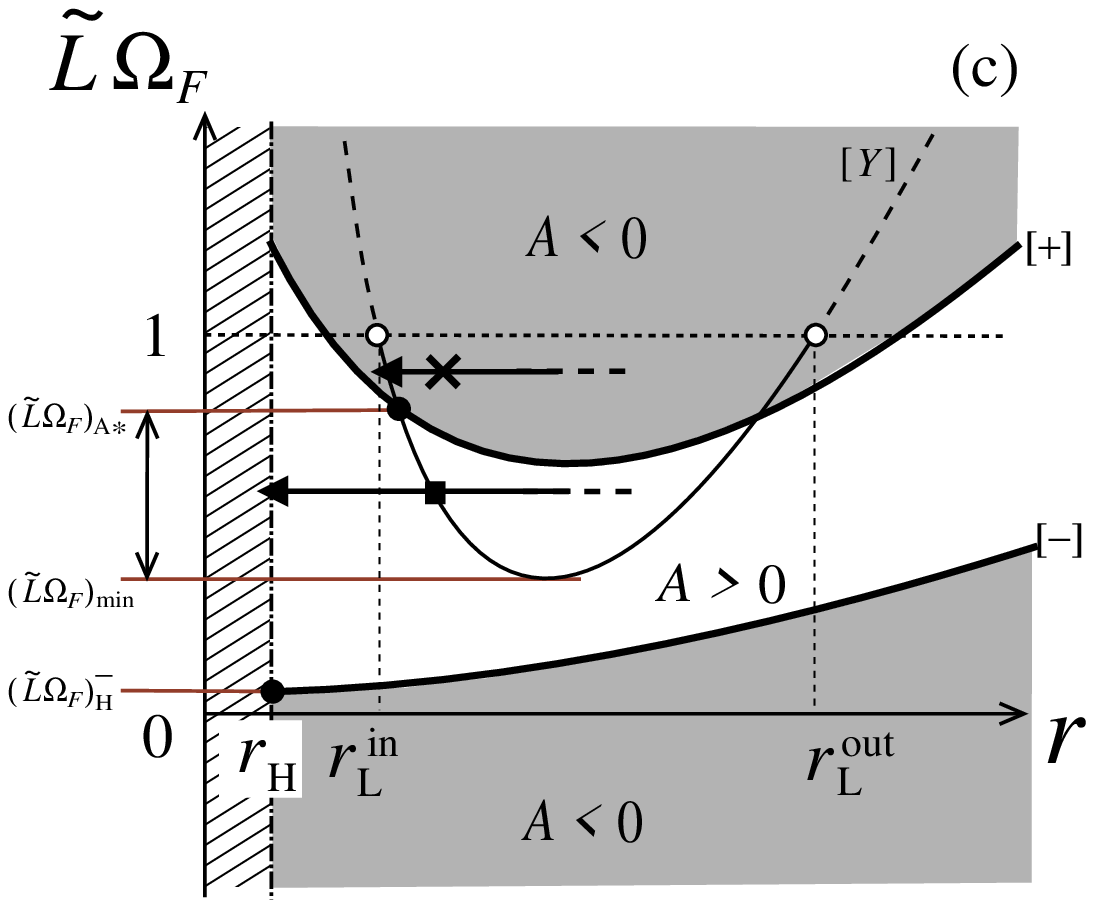} \hspace{0.5cm}
  \includegraphics[scale=0.75]{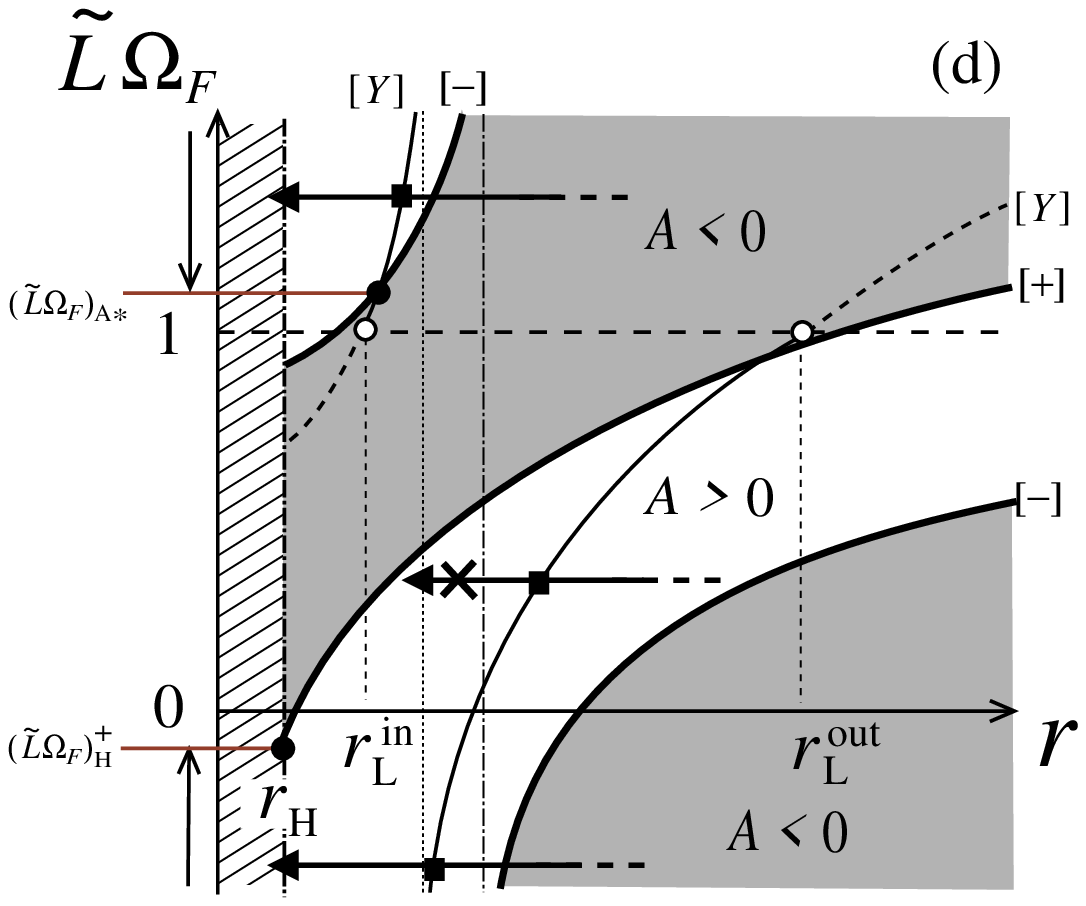}
 \caption{
  (TOP) 
   The possible range of $\Lt\Omega_F$ for (a) type I and (b) type II
   cases. The vertical axis is $\Lt\Omega_F$, while the horizontal axis
   is the Alfv\'{e}n radius.  Curves $\Lt\Omega_F = Y(r_{\rm A})$
   (marked by $[Y_{\rm A}]$) and $k(r_{\rm A}) = 0$ (marked by 
    $[k_{\rm A}]$) are plotted. Two crossing points between the 
   $\Lt\Omega_F = Y(r_{\rm A})$ curve and $\Lt\Omega_F =$ constant line 
   in the type I case ({\it left\/}) show the inner and outer Alfv\'{e}n
   radii. On the other hand, one Alfv\'en radius exists in the type II
   case ({\it right\/}). 
   The hatched region bounded by $k_{\rm A}=0$ shows the forbidden
   region for the parameter.    
   For type I case the range of $(\Lt\Omega_F)_{\rm min} < \Lt\Omega_F <
   1$ is acceptable as a trans-Alfv\'enic solution, while for
   type II case the ranges of $\Lt\Omega_F < (\Lt\Omega_F)_{\rm max}$
   and $ 1 < \Lt\Omega_F$ are acceptable.  
   (BOTTOM) 
   Curves $\Lt\Omega_F = (\Lt\Omega_F)^\pm$ (marked by $[+]$ and $[-]$)
   and  $\Lt\Omega_F = Y$ (marked by $[Y]$) for (c) type I and (d) 
   type II cases.  Some accretion solutions with certain $\Lt\Omega_F$
   values are shown by horizontal arrows, where the crossing point with
   the $\Lt\Omega_F = Y$  curve (i.e., $Y=Y_{\rm A}$) indicates the
   Alfv\'{e}n point (marked by the fill square).  When the solution
   crosses the $\Lt\Omega_F = (\Lt\Omega_F)^\pm$ curve, where $A=0$,
   the Mach number of the solution diverges there; that is, no physical 
   accretion solution exits (the horizontal arrow marked by the cross).  
   The shaded regions show the $A < 0$ regions. 
   Note that the $k > 0$ region is enclosed in the $A < 0$ region. 
   In these schematic diagrams, we set up the field configuration
   $\xi^2$ by equation (\ref{eq:xi_Iin}) for type~I and equation
  (\ref{eq:xi_IIin}) for type~II near the equator as typical models. 
 }
 \label{fig:A-region}
\end{figure*}

 Now, we consider magnetic field line connected to a black hole with a
 certain value of $\Omega_F$.  Figures~\ref{fig:A-region}a
 and~\ref{fig:A-region}b show the locations of the Alfv\'{e}n radii for
 a given $\Lt\Omega_F$ value.  When the magnetic field line rotates
 faster than the black hole (i.e., a slowly rotating black hole case), 
 $0 \leq \omega_{H} \leq \Omega_F$, whose state is named ``type I'' in
 Takahashi et al.\cite{TNTT90}, the condition for $\Lt\Omega_F$ for the
 existence of the Alfv\'{e}n point in the flow solution is
 $(\Lt\Omega_F)_{\rm min} < \Lt\Omega_F \leq 1$  (see
 Fig.\ref{fig:A-region}a), where $(\Lt\Omega_F)_{\rm min}$ is the
 minimum value of $\Lt\Omega_F$ for the Alfv\'{e}n point to appear on  
 the flow, and it is given by $d Y_{\rm A}/d r_{\rm A}=0$. The MHD flow
 with $E>0$ and $L>0$ only is obtained.

 For the type I case, it is possible to select the inner or outer
 Alfv\'{e}n point in an accretion solution, and then trans-Alfv\'{e}nic
 accretion solutions with two types are acceptable (see
 \cite{Takahashi02}); that is, ``magneto-like'' and ``hydro-like''
 accretion solutions. 
 The magneto-like accretion solution passes through the {\it inner}\/
 Alfv\'en point and the inner fast magnetosonic point, and the
 hydro-like accretion solution passes through the {\it outer}\/ Alfv\'en
 point and the middle fast magnetosonic point.  Note that, in the latter
 type solution case, the acceptable range of $\Lt\Omega_F$ is modified
 to $(\Lt\Omega_F)_{\rm min} < \Lt\Omega_F\leq (\Lt\Omega_F)_{\rm max}$,
 where $ (\Lt\Omega_F)_{\rm max}$ is the maximum value of $\Lt\Omega_F$
 to avoid the forbidden region discussed in \cite{TNTT90} 
 between the outer Alfv\'{e}n point and the event horizon, and is given
 by $d k_{\rm A}/d r_{\rm A}=0$ with $k_{\rm A}=0$.   
 For a counterrotating case $a\Omega_F < 0 $ (named ``type III''), we
 see the similar restrictions on $\Lt\Omega_F$ as type I, although we
 obtain $L<0$ and $E>0$ flows.   

 Figure~\ref{fig:A-region}b shows the rapidly rotating black hole case of 
 $0 <\Omega_F <\omega_{H}$ (named ``type II''). For a certain
 $\Lt\Omega_F$ value, one Alfv\'en point is obtained on the flow
 solution. In this case, the negative energy accretion with $E < 0$ and
 $L < 0$ is possible when $\Lt\Omega_F \geq 1$.  The flow with $E > 0$ 
 and $L\leq 0$ is realized when $\Lt\Omega_F\leq 0$.  
 For $0 < \Lt\Omega_F\leq (\Lt\Omega_F)_{\rm max}$, the inflow with 
 $E > 0$ and $L > 0$ is realized.  More detail discussions about the
 restriction by the Alfv\'{e}n points are presented by Takahashi et
 al.\cite{TNTT90}.

\subsection{ Super-Alfv\'{e}nic Flow }        \label{sec:super-fast}

 After passing through the Alfv\'{e}n point, the ingoing/outgoing flow
 solution passes through the fast magnetosonic point automatically.
 However, if the location of $A(r;\Psi)=0$ appears on the flow solution 
 considered in the super-Alfv\'{e}nic region, the Mach number of this
 super-fast magnetosonic flow solution diverges there; that is, no
 physical MHD accretion/wind solution exists.  To obtain a physical
 MHD accretion/wind solution, no $A = 0$ surfaces in the
 super-Alfv\'{e}nic region must be required.   
 Although the value of $A$ depends on the flow parameters $\Omega_F$,
 $E$ and $L$ under a given function $\xi(r,\theta)$ and the hole's spin 
 $a$, we will find restrictions on these combinations of parameters to
 avoid the appearance of the $A = 0$ location.

 The condition $ A > 0 $ for a super-Alfv\'{e}nic flow can be reduced to   
\begin{widetext}
 \begin{equation}
    A(r; \Psi) =   \frac{ -G_t (1+Y+X) \EE^2 }{ (G_\phi \Omega_F)^2 } 
   \left[ \Lt\Omega_F - (\Lt\Omega_F)^{+} \right]
   \left[ \Lt\Omega_F - (\Lt\Omega_F)^{-} \right] > 0   \label{eq:A=0}
 \end{equation}
  with 
 \begin{equation}
   (\Lt\Omega_F)^\pm \equiv \frac{ Y }{ 1+Y+X } \left\{ (1+X) \pm 
    \epsilon 
    \left[ 1 + (1-Y)X - (1+Y+X)\frac{G_t}{\hat{E^2}} \right]^{1/2}
    \right\} \ ,                        \label{eq:c-jj}
 \end{equation}
\end{widetext}
 where $ X \equiv g_{\phi\phi}G_t (1-\xi^2)/\rho_w^2$ and $\epsilon
 \equiv \Omega_F (\Omega_F-\omega) / | \Omega_F (\Omega_F-\omega) |$.    
 When the function $\xi^2 = \xi^2(r; \Psi)$ is specified as a given
 magnetic field structure with the field aligned flow parameters
 $\Omega_F$ and $\EE$, the value of $(\Lt\Omega_F)^\pm$ is determined
 along the stream line.  
 In Figures~\ref{fig:A-region}c and ~\ref{fig:A-region}d we show the 
 $\Lt\Omega_F=(\Lt\Omega_F)^\pm$ curves for the slowly and rapidly
 rotating black hole cases, respectively. The models of $\xi^2(r; \Psi)$
 are discussed in Appendix~\ref{sec:beta}.

 To obtain a trans-fast magnetosonic accreting flow onto a black hole,
 the $\Lt\Omega_F=$ constant line must not cross the 
 $\Lt\Omega_F = (\Lt\Omega_F)^\pm$ curves between the Alfv\'en point and
 the event horizon on the $\Lt\Omega_F$ -- $r$ plane.  The condition
 $A>0$ for the super-Alfv\'{e}nic flow solution is satisfied when 
\begin{description}
 \item[(i)]  $(\Lt\Omega_F)^{-} < \Lt\Omega_F < (\Lt\Omega_F)^{+}$ 
\end{description}
 in the $G_t (1+Y+X) > 0$  region, and  
\begin{description}
 \item[(ii)] $\Lt\Omega_F < (\Lt\Omega_F)^{+}$ or  $\Lt\Omega_F >
              (\Lt\Omega_F)^{-}$ 
\end{description}
 in the $G_t (1+Y+X) < 0$ region.  
 The value of $(\Lt\Omega_F)^{-}$ diverges at the location of
 $X=-(1+Y)$, while $(\Lt\Omega_F)^{+}$ has a non-zero finite value there.  
 Thus, the value of $\Lt\Omega_F$ (or the location of the Alfv\'{e}n
 point) is restricted. 
 Examples of this restriction on $\Lt\Omega_F$ for the $A>0$ regions are
 also shown in Figures~\ref{fig:A-region}c and~\ref{fig:A-region}d for
 Type~I and~II cases, respectively.  
 For accretion onto a slowly rotating black hole (type I; see
 Fig.~\ref{fig:A-region}c),   
 we see that the case (i) is applied everywhere.  
 For type~I and~III, the location of $X=-(1+Y)$ may exist, although it
 depends on a model of $\xi^2$; that is, the case (ii) may arise. In
 this case, however, there is no additional restriction in the
 $(\Lt\Omega_F)_{\rm min} < \Lt\Omega_F < 1$ range.  On the other hand, 
 the case (ii) arises near the event horizon for a rapidly rotating
 black hole case (type II; see Fig.~\ref{fig:A-region}d ).   
 Note that for type~II the case (i) is also possible just outside the
 event horizon, although it depends on $\xi$ (to be discussed in
 \S~\ref{sec:inc}). In this case, the negative energy inflow solutions
 only are obtained; the positive energy inflow solutions are forbidden
 because the $A<0$ region appears on the way to the horizon. 


 In the last of this section, we should mention that there is the
 innermost limit of the inner Alfv\'en point (labeled by ``A$\ast$'' in
 Figs.~\ref{fig:A-region}c and~\ref{fig:A-region}d) under the given
 parameters $\Omega_F$ and $\hat E$.  This limit gives the restriction
 on the $\tilde{L}\Omega_F$ value, which is the maximum value for type~I
 case and the minimum value for type~II case.  The details are discussed
 in Appendix \ref{sec:A-star}.

\section{ Boundary Conditions at the Event Horizon }  \label{sec:bc-EH}

\subsection{ Inclination of Magnetic Field Lines }     \label{sec:inc}

 Without the trans-field equation and the informations for the plasma
 sources (as the initial conditions for MHD flows), we can not 
 obtain the concrete function of $\xi(r,\theta)$ in the black hole
 magnetosphere considered.  Nevertheless, just on the event horizon, 
 we have already known that $\xi_{H}=1$ (or $\beta_{H}^2=-\alpha_{H}$),  
 which is the boundary condition at the horizon.  However, just outside 
 the event horizon, the magnetic field configuration depends on the
 plasma inertia effects.  Here, we discuss the function
 $\xi^2(r,\theta)$ near the event horizon.   
 One may expect that the signature of the function
 $A_{H}(\theta) \equiv A(r_{H},\theta)$ is at least determined for any
 field aligned flow parameter sets.  However, at the event horizon, the
 value of the function $X$ in equation (\ref{eq:A=0}) can not be
 specified by only the condition $\xi_{H}^2=1$.  
 Now, we will expand the function $\xi^2$ as $\xi^2 = 1+\chi (r/r_{H}-1)$, 
 where $\chi\equiv r_{H}(\partial_r\xi)_{H}$ represents the magnetic
 field configuration near the horizon. To obtain physical MHD accretion
 solutions satisfying the condition $A_{H}>0$, we should discuss the
 restrictions on the value of $\chi$ and on the allowable ranges of 
 the field aligned flow parameters (e.g., $E$, $L$ and $\Omega_F$),
 which should be consistent with the boundary condition at the plasma
 sources.  Note that the function $\chi$ can be expressed as the
 differentials of  $\Bp^2$ and $\Bf^2$ at the event horizon.

 The restrictions on the function $\chi(\theta)$ specify how the
 inclination angles change (having poloidal and toroidal components)
 going away from the horizon.  In other words, the $\chi$ value
 represents the deviation of the inclination (or pitch) angle at the
 horizon from the $\xi^2(r,\theta) = 1$ model.  To consider this
 situation, we also expand the magnetic field components as 
 ${\cal B}_p = {\cal B}_{p H} + \delta{\cal B}_{p H} (r/r_H - 1)$ and   
 ${\cal B}_\phi = {\cal B}_{\phi H} + \delta{\cal B}_{\phi H} 
  (r/r_H - 1)$.  Then, we obtain
 \begin{equation}
    \left( \frac{{\cal B}_p}{{\cal B}_\phi} \right)^2 = 
    \frac{1}{(-\alpha_{H})} 
    \left[ 1 + 2 \left( \frac{\delta{\cal B}_p}{{\cal B}_p} 
             - \frac{\delta{\cal B}_\phi}{{\cal B}_\phi} \right)_{H}
               \left(\frac{r}{r_{H}} -1 \right) \right] \ . 
 \end{equation}
 When we modify the magnetic field configuration  
 $(\delta{\cal B}_p/{\cal B}_p)_{H} > 0$ and/or  
 $(\delta{\cal B}_\phi/{\cal B}_\phi)_{H} < 0$, we see that the value of
 $\chi$ increases.  Although the term ${\cal B}_p$ means the divergence
 of the magnetic flux tube to the ($+r$)-direction, when the ratio of
 the divergence $(\delta{\cal B}_p/{\cal B}_p)_{H}$ for a certain model
 is larger than that of the $\xi^2=1$ ($\chi=0$) model, we see that 
 $(\delta{\cal B}_p/{\cal B}_p)_{H} > 0$.  Furthermore, if the bending
 angle to the toroidal direction is smaller than that in the force-free
 case ($\delta{\cal B}_\phi$=0 along a magnetic field line), we see that  
 $(\delta{\cal B}_\phi/{\cal B}_\phi)_{H} < 0$.

\begin{figure*}
   \includegraphics[scale=0.48]{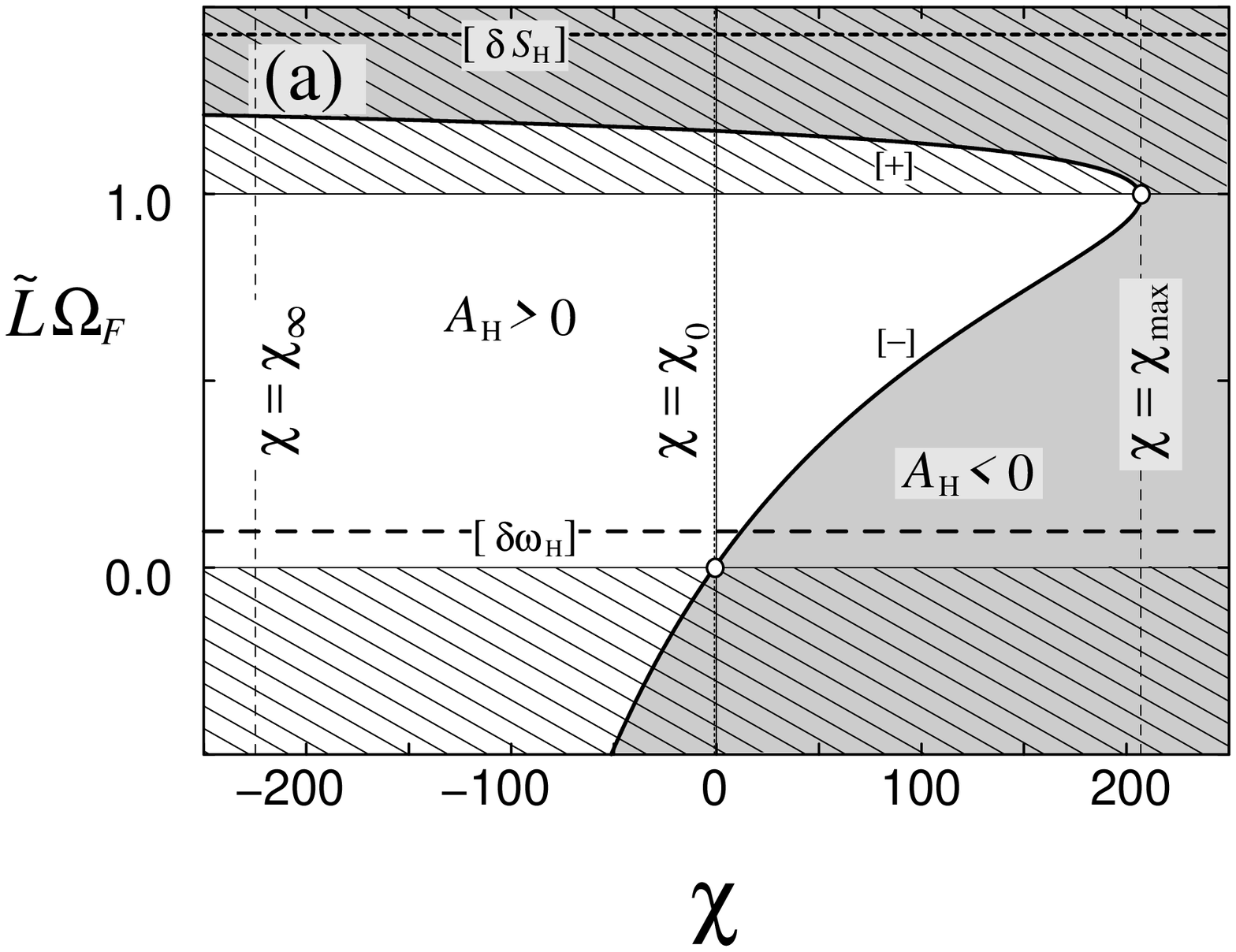} \hspace{0.5cm}
   \includegraphics[scale=0.48]{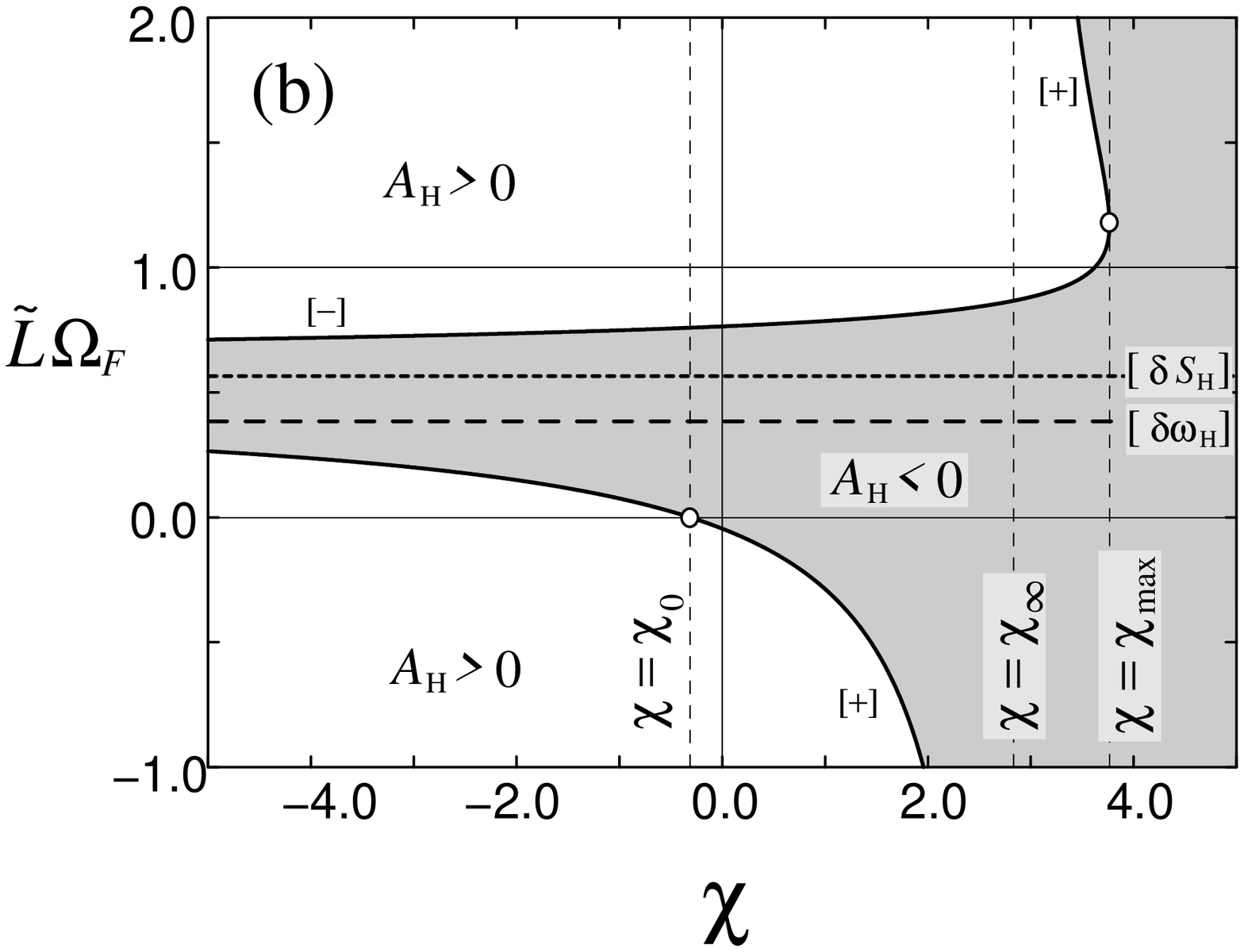}
  \caption{
   The horizon's boundary condition on $\Lt\Omega_F$ for (a) type~I of
   $a=0.3m$ and (b) type~II of $a=0.9m$.  The bold thick curves denote
   the $\Lt\Omega_F=(\Lt\Omega_F)_{H}^\pm (\chi)$ curves (marked $[+]$ and
   $[-]$ respectively) with $\EE^2=1.0$, 
   $\Omega_F=0.5\Omega_{\rm max}=0.09623/m$ and $\theta=\pi/2$.  
   The sheaded region corresponds to the $A_{H}<0$ region, which is
   forbidden as a black hole accretion solution.  The hatched regions 
   in (a) show the $\Lt\Omega_F>1$ and $\Lt\Omega_F<0$ regions
   (unphysical).   
   The $\Lt\Omega_F = Y_{H}$ line is denoted by the dotted line and
   labeled by [$\delta S_{H}$]. 
   The horizontal broken line marked by [$\delta\omega_H$] shows
   the $\Lt\Omega_F=(\Lt\Omega_F)_\omega$ line.  
   Although, in (b), it seems $(\Lt\Omega_F)_{H}^{+} <
   (\Lt\Omega_F)_\omega < (\Lt\Omega_F)_{H}^{-}$ in this plot, we can
   find the case $(\Lt\Omega_F)_\omega < (\Lt\Omega_F)_{H}^{+} < Y_{H}$
   for larger $(-\chi)$ values. For the range $(\Lt\Omega_F)_\omega <
   \Lt\Omega_F < (\Lt\Omega_F)_{H}^{+}$, the spin-up is caused for
   type~II.    
   }
 \label{fig:Y_H}
\end{figure*} 

 If the location of $A=0$ appears between the Alfv\'{e}n point and the
 event horizon (i.e., the super-Alfv\`enic region of the flow), such 
 a solution is unphysical as an accretion solution as mentioned in
 \S~\ref{sec:super-fast}.  At the event horizon, to accrete onto the
 black hole, the condition $A_{H}(\chi)>0$ must be required.   
 The boundaries of the $A_{H}(\chi)>0$ region on the 
 $\chi$--$\Lt\Omega_F$ plane are obtained by 
\begin{widetext}
 \begin{equation}
    (\Lt\Omega_F)_{H}^\pm (\chi) 
       =  \frac{ \Omega_F }{ \Omega_F +\omega_{H} +\omega_{H} X_{H} }
          \left\{ ( 1 + X_{H} )  
         \pm \, \epsilon_{H} \left[ 1 + 
         \left( \frac{\Omega_F^2-\omega_{H}^2}{\omega_{H}^2} \right)
         \frac{g_{tt}^{H}}{\EE^2} -  
         \left( \frac{\Omega_F-\omega_{H}}{\omega_{H}} \right)
         \left(1-\frac{g_{tt}^{H}}{\EE^2}\right) X_{H} \right]^{1/2} 
         \right\} \ ,  
     \label{eq:lt-H}
 \end{equation}
\end{widetext}
 where 
 $\epsilon_{H} \equiv \Omega_F (\Omega_F-\omega_{H}) / | \Omega_F
 (\Omega_F-\omega_{H}) |$, $X_{H} \equiv X(r_{H}) = {\cal H} \chi$ and   
 ${\cal H} \equiv 2m^2 r_{H} G_t^{H}/[(r_{H}-m)\Sigma_{H}]$.  
 Figure~\ref{fig:Y_H} shows the relation between $(\Lt\Omega_F)^\pm_{H}$
 and $\chi$ for type~I and~II cases.  There is a maximum $\chi$ value
 ($\equiv \chi_{\rm max}$) for existing $A>0$ region in both type~I
 and~II cases. 


 The value of $(\tilde{L}\Omega_F)_{H}^{-}$ becomes zero at
 $\chi=\chi_0$, and diverges at $\chi=\chi_\infty$. Although these
 characteristic $\chi$'s depend on the flow parameters, the details are
 discussed in Appendix \ref{sec:chi}.  Furthermore, the relations
 between the $\chi$ value and the acceptable ranges of
 $\tilde{L}\Omega_F$ are also summarized there.

\subsection{ Black Hole Spin-Up/Down via MHD Accretion }  \label{sec:spin}

 In \S~\ref{sec:super-fast} and \S~\ref{sec:inc} (see also Appendix
 \ref{sec:chi}), we have discussed the restriction on $L/E$ for the MHD
 accretion onto a black hole to avoid the $A=0$ surface on the flow
 solution, by considering the conditions at both the event horizon and
 the Alfv\'en point.    
 Here, we will discuss the increase or decrease of the angular velocity
 (spin-up/down) of a rotating black hole by applying this restriction on
 $L/E$ to equation (\ref{eq:spin-BH}).  Hereafter, we treat the case of
 $a > 0$ and $\Omega_F > 0$ mainly (i.e., type~I and~II cases).

 Although the accreting gas carries the mass and angular momentum 
 into the black hole, 
 both the cases $\delta\omega_{H}/\omega_{H} > 0$ (spin-up) and 
 $\delta\omega_{H}/\omega_{H} < 0$ (spin-down) are possible for the
 positive energy $(E>0)$ MHD inflows.  That is, the range of 
 $\Lt\Omega_F > (\Lt\Omega_F)_{\omega}$ gives the hole's spin-up and the 
 range of $\Lt\Omega_F < (\Lt\Omega_F)_{\omega}$ gives the spin-down,   
 where 
 \begin{equation}
    (\Lt\Omega_F)_{\omega}\equiv\left( 1+ r_{H}/m \right)a\Omega_F ~;   
 \end{equation}
 in Figure~\ref{fig:Y_H} we show the $\Lt\Omega_F=(\Lt\Omega_F)_\omega$
 line (where $\delta\omega_{H}=0$).       
 [Similarly, with respect to the spin parameter $a$, we see that $\delta
 (a/m) > 0$ (or $\delta (a/m) < 0$) for positive energy MHD inflows
 with $\Lt\Omega_F > 2a\Omega_F$ (or $\Lt\Omega_F < 2a\Omega_F$).]  
 Furthermore, by considering the restrictions from the Alfv\'en point
 and the event horizon, we obtain that increase of the angular velocity
 of the black hole $\delta\omega_{H}>0$ by MHD inflows is realized when  
 $(\Lt\Omega_F)_\omega < \Lt\Omega_F < (\Lt\Omega_F)_{\rm A \ast}$ 
 for type~I case, and 
 $(\Lt\Omega_F)_\omega < \Lt\Omega_F < (\Lt\Omega_F)_{H}^{+}$  
 for type~II case. 
 Note that, even if positive angular momentum $L>0$ of MHD flow falls
 into the black hole, the angular velocity of the black hole can be
 decreased; that is, $\delta\omega_{H} < 0$ is obtained for 
 $0 < \Lt\Omega_F < (\Lt\Omega_F)_{\omega}$.

 For a slowly rotating black hole of type I case, equatorial
 (low-latitude) inflows onto the black hole contribute to the spin-up,
 but spin-down effects can be also realized by MHD plasma accreting 
 from the polar (higher latitude) region of the event horizon. 
 The value of $(\Lt\Omega_F)_{\rm min}$ can be specified by the
 latitude of the Alfv\'{e}n point, where $r_{\rm A}(\theta_{\rm A}) =
 r_{\rm A}^{\rm in} = r_{\rm A}^{\rm out}$, so that we define the
 critical angle of $\theta_{\rm A} ~[\equiv (\theta_{\rm A})_{\rm cr}]$
 by $(\Lt\Omega_F)_{\rm min}=(\Lt\Omega_F)_\omega$, except for a
 Schwarzschild black hole, where $(\Lt\Omega_F)_\omega$  becomes zero.  
 When $\theta_{\rm A} > (\theta_{\rm A})_{\rm cr}$ along a flow
 considered, the relation of $\Lt\Omega_F = (\Lt\Omega_F)_{\rm min} >
 (\Lt\Omega_F)_\omega$ is satisfied and the MHD inflows only contribute
 to the spin-up of the black hole.   
 On the other hand, when $\theta_{\rm A} < (\theta_{\rm A})_{\rm cr}$,
 the situation $(\Lt\Omega_F)_{\rm min} < \Lt\Omega_F <
 (\Lt\Omega_F)_\omega$ (the black hole's spin-down) is also possible;
 note that, even if $\theta_{\rm A}<(\theta_{\rm A})_{\rm cr}$, 
 the MHD inflows in the range of $(\Lt\Omega_F)_\omega < \Lt\Omega_F <
 (\Lt\Omega_F)_{\rm A \ast}$ can lead the black hole into the spin-up.  
 For larger values of $a$ or for smaller values of $\Omega_F$, such a
 spin-down range specified by the critical value angle 
 $(\theta_{\rm A})_{\rm cr}$ expands toward lower latitude regions,
 while the value of $(\tilde{L}\Omega_F)_\omega$ increases; in the 
 $a\to M$ limit, we see that $(\tilde{L}\Omega_F)_\omega \to Y_{H} =
 \Omega_F/\omega_{H}$ ($>1$). Thus, the spin-down effect dominates 
 for a rapidly rotating black hole of type~I; that is, even if 
 $\Omega_F > \omega_{H}$, the MHD inflows spin down the hole's rotation.

 For a rapidly rotating black hole of type II case, both $L>0$ and $L<0$
 inflows are possible.  The inflow with $L>0$ decreases the spin of the
 black hole when $0 < \tilde{L}\Omega_F < (\tilde{L}\Omega_F)_\omega$.  
 On the other hand, the $L>0$ inflows increase the hole's spin when
 $(\tilde{L}\Omega_F)_\omega < \tilde{L}\Omega_F <
 (\tilde{L}\Omega_F)_{H}^{+}$. However, after the black hole spin-up,
 the value of $(\tilde{L}\Omega_F)_\omega$ increases, so that the
 parameter range of $\tilde{L}\Omega_F$ for the spin-up decreases. 
 Note that, in the $a\to m$ limit, any ingoing flows with $L>0$ cannot
 increase the angular velocity and the spin parameter of the black hole
 no more, because both $(\Lt\Omega_F)_\omega$ and $2a\Omega_F$ become
 $Y_{H}$, where $(\tilde{L}\Omega_F)_{H}^{+} < Y_{H} <
 (\tilde{L}\Omega_F)_{H}^{-}$; that is the $A_{H}>0$ region. 
 On the other hand, in type~II case, the negative angular momentum MHD
 inflows are also possible. In such a case, we always obtain 
 $\delta \omega_{H}<0$  and $\delta(a/m) < 0$ as expected.   
 Furthermore, if we consider the negative energy ($E<0$) MHD inflows
 \cite{TNTT90}, 
 which always carry the negative angular momentum ($L<0$), we always see
 that $\delta\omega_{H}/\omega_{H} < 0$  and $\delta (a/m) < 0$.

\begin{figure*}[t] 
   \includegraphics[scale=0.97]{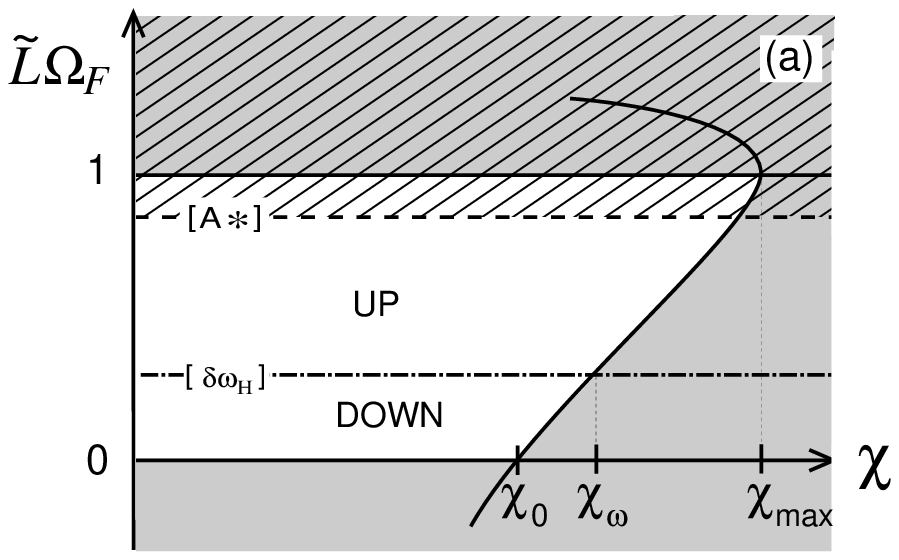} 
   \includegraphics[scale=0.97]{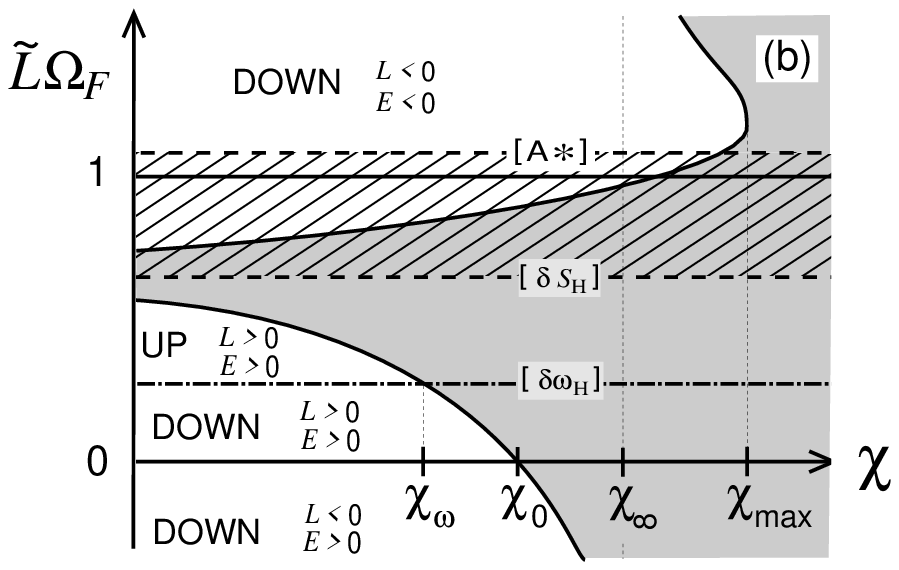}
 \caption{
   Summary of the $\Lt\Omega_F$ ranges for the spin-up/down of a black 
   hole in (a) type~I case and (b) type~II case.  The value of
   $\Lt\Omega_F$ for acceptable MHD accretion is restricted by at least
   the boundary condition at the event horizon ($A_{H}>0$) and the
   condition of the Alfv\'en point for existing in the $A>0$ region. 
   The shaded forbidden region is the same as that of Fig.~\ref{fig:Y_H}. 
   The labels [A$\ast$], [$\delta\omega_{H}$] and [$\delta S_{H}$] show
   the $\Lt\Omega_F=(\Lt\Omega_F)_{A\ast}$ line, the
   $\Lt\Omega_F=(\Lt\Omega_F)_{\omega}$ line and the $\Lt\Omega_F=Y_{H}$
   line, respectively.  The hatched region is forbidden by the condition
   at the Alfv\'en  point. 
    For type~I case, the range of $(\Lt\Omega_F)_{\rm min} < \Lt\Omega_F
    < (\Lt\Omega_F)_{\rm A \ast}$ is acceptable, and the MHD inflows
    with $\theta_{\rm A} > (\theta_{\rm A})_{\rm cr}$ always cause the
    hole's spin-up.        
 }
 \label{fig:spin}
\end{figure*} 

 Figure~\ref{fig:spin} shows the summary of the hole's spin-up/down by
 MHD inflows, which also includes the contribution by the magnetic field
 parameter $\chi$.  First, in the case of type~I (see
 Fig.~\ref{fig:spin}a), the spin-up of $\delta\omega_{H}/\omega_{H} > 0$
 is obtained when the value of $\tilde{L}\Omega_F$ is in the range of
 ${\rm MAX}[(\Lt\Omega_F)_{\rm min}, (\Lt\Omega_F)_\omega,
 (\Lt\Omega_F)_{H}^{-}] < \Lt\Omega_F < (\Lt\Omega_F)_{\rm A \ast}<1 $.   
 The spin-down of $\delta\omega_{H}/\omega_{H} < 0$ is obtained     
 when ${\rm MAX}[ (\Lt\Omega_F)_{\rm min}, (\Lt\Omega_F)_{H}^{-} ] < 
 \Lt\Omega_F < (\Lt\Omega_F)_\omega$, which is only realized for inflows
 with $\theta_{\rm A}<(\theta_{\rm A})_{\rm cr}$.    
 Next, for type~II case (see Fig.~\ref{fig:spin}b), we obtain the
 spin-up of $\delta\omega_{H}/\omega_{H} > 0$ when 
 $(\Lt\Omega_F)_\omega < \Lt\Omega_F < (\Lt\Omega_F)_{H}^{+}$ for 
 $\chi < \chi_\omega$, where $\chi_\omega$ is defined by the condition
 of $(\Lt\Omega_F)_{H}^{+} = (\Lt\Omega_F)_\omega$. 
 The spin-down of $\delta\omega_{H}/\omega_{H} < 0$ is obtained when (1) 
 $\Lt\Omega_F < {\rm MIN}[(\Lt\Omega_F)_\omega, (\Lt\Omega_F)_{H}^{+}]$
 or $(\Lt\Omega_F)_{\rm A \ast} < \Lt\Omega_F$ for $\chi < \chi_\infty$  
 and (2) $(\Lt\Omega_F)_{\rm A \ast} < \Lt\Omega_F <
 (\Lt\Omega_F)_{H}^{+}$ for $\chi_\infty < \chi < \chi_{\rm max}$. 
 When we consider a counterrotating magnetosphere of $\Omega_F<0$
 (type~III), where $a>0$ is considered, we have the relation
 $\Lt\Omega_F > 0 > 2a\Omega_F > (\Lt\Omega_F)_\omega$. 
 In this type~III case, both the angular momentum and the spin parameter
 of the black hole decrease; that is, $\delta \omega_{H}/\omega_{H} < 0$
 and $\delta(a/m) < 0$.

 From the definition of the surface area of the black hole 
 $S_{H} \equiv 4\pi (r_{H}^2 + a^2)$, we obtain the relation
 $\delta S_{H} \propto E - \omega_{H}L$.  Here, we should remember
 that, in the case of type~II, the acceptable range of
 $\tilde{L}\Omega_F$ for positive energy ($E>0$) MHD inflows is given as 
 $\Lt\Omega_F < (\Lt\Omega_F)_{H}^{+} \leq Y_{H}$, while $\Lt\Omega_F <
 1 < Y_{H}$ in the case of type~I and 
 $\Lt\Omega_F > 0 > Y_{H}$ in the case of type~III.  So, for $E>0$ flows
 (in all cases of types~I,~II and~III),  we can show $\delta S_{H}>0$,
 which means the increasing surface area of a  black hole.    
 On the other hand, for $E<0$ flows (that can be possible in the case of
 type~II), the relation of $Y_{H} < \Lt\Omega$ is always satisfied, 
 and then the relation of $\delta S_{H}>0$ is also satisfied.  Thus, the 
 area low $\delta S_{H}>0$ of the black hole is confirmed for stationary
 and axisymmetric ideal MHD accretion flows.

\subsection{ Secular Evolution of the Black Hole Spin }  \label{sec:evo}

 In \S~\ref{sec:spin}, we mentioned that the MHD inflow prevents the
 evolution to the maximally rotating black hole.  This is because
 $\delta\omega_H < 0$ and $\delta(a/m) < 0$ for the extreme rotating
 black hole limit.  On the other hand, we know that $\delta\omega_H>0$
 and $\delta(a/m)>0$ for the non-rotating black hole case.  Now, we will
 discuss a secular evolution of the black hole spin by MHD accretion. 
 Here, we assume that the values of $L(\Psi)$, $E(\Psi)$ and
 $\Omega_F(\Psi)$ keep constants for a magnetic flux tube considered 
 (a $\Psi$=constant line) during the evolution. 
 Furthermore, we do not discuss the plausible configuration of the
 magnetic field as a global solution of the black hole magnetosphere.

 First, we consider the case of type~I ($0 < \omega_{H} < \Omega_F$).
 When the magnetized plasma of $(\Lt\Omega_F)_\omega < \Lt\Omega_F <
 (\Lt\Omega_F)_{\rm A \ast}$ falls into the black hole, the spin of the
 black hole will increase in secular time-scale.  Then, the value of
 $(\Lt\Omega_F)_\omega$ also increases, and the state of
 $\tilde{L}\Omega_F = (\tilde{L}\Omega_F)_\omega$, where
 $\delta\omega_{H} = 0$, will be realized sometime.  In this state, the 
 evolution of the black hole's spin is terminated.   
 On the other hand, when $(\Lt\Omega_F)_{\rm min} < \Lt\Omega_F <
 (\Lt\Omega_F)_\omega$, the black hole's spin decreases, and the value
 of $(\Lt\Omega_F)_\omega$ is also decreases.  The final state will be  
 settled to $\tilde{L}\Omega_F = (\tilde{L}\Omega_F)_\omega$. 
 Although the spin value $a$ terminated by accreting MHD plasma is given
 by $(\tilde{L}\Omega_F)_\omega = \tilde{L}\Omega_F$, the values of $L$,
 $E$ and $\Omega_F$ should be specified by a magnetized accretion disk
 model.  
 Note that, during the spin-up stage, the situation of 
 $\omega_{H} \to \Omega_F$ may be realized before the $\tilde{L}\Omega_F =
 (\tilde{L}\Omega_F)_\omega$ state.   
 In this case, the spin value $a$ terminated is specified by
 $\Omega_F$.

 Next, we consider the type~II case ($0 < \Omega_F < \omega_{H}$).  When
 magnetized plasma with $(\tilde{L}\Omega_F)_\omega < \tilde{L}\Omega_F
 < (\tilde{L}\Omega_F)_{H}^{+}$, the black hole will spin up in secular
 time scale. In this case, the value of $(\tilde{L}\Omega_F)_\omega$
 also increases and the situation of $\tilde{L}\Omega_F =
 (\tilde{L}\Omega_F)_\omega$ would be realized. 
 When $0 < \tilde{L}\Omega_F < (\tilde{L}\Omega_F)_\omega$, the black
 hole spins down in spite of the positive angular momentum inflows. 
 With decreasing the hole's spin, the value of
 $(\tilde{L}\Omega_F)_\omega$ also decreases, and the state of
 $\tilde{L}\Omega_F = (\tilde{L}\Omega_F)_\omega$ would be realized.
 One may expect that, by decreasing the angular velocity of the black
 hole, the state of $\omega_{H} \to \Omega_F$ is achieved before
 reaching the state of $\tilde{L}\Omega_F = (\tilde{L}\Omega_F)_\omega$.   
 However, the state of $\omega_{H} \to \Omega_F$ is only possible for
 the $L < 0$ inflows (see Appendix \ref{sec:chi}).  So that, the inflows 
 with $L > 0$ are terminated to the state of $\tilde{L}\Omega_F =
 (\tilde{L}\Omega_F)_\omega$. 
 In type~II case, the negative angular momentum inflows are possible
 when $\tilde{L}\Omega_F < 0$ or $\tilde{L}\Omega_F >
 (\tilde{L}\Omega_F)_{A \ast}$.  The rotating black hole spins down, and
 reaches the state $\omega_{H} \to \Omega_F$.  For the magnetically
 dominated accretion flows, where $E<0$ and $L<0$, such a  state would
 be only realized after a long time spin evolution.  
 Although the state $\omega_{H} \to \Omega_F$ can be achieved in both
 type~I and~II, the transition from type~I to type~II or the reverse is
 impossible.  This is because the asymptotic evolution
 $\omega_{H}\to\Omega_F$ is possible for $L>0$ inflows of type~I or
 for $L<0$ inflows of type~II.  The transition between type~I and~II 
 contradicts the model's assumption that the field-aligned parameters do
 not change during the hole's evolution.

 In this section, we have discussed about the spin-up/down effects 
 along {\it one}\/ magnetic flux tube.  However, in actual black hole
 magnetosphere models, the above arguments could be considered for 
 each magnetic flux, and the effects on the black hole spin should be
 integrated over the whole magnetic fluxes across the event horizon.  
 In type~I case, although the equatorial inflow would dominate in the
 black hole accretion and such an inflow contributes to the spin-up, but
 the spin-down by the inflow in the polar region may not be negligible
 in a black hole magnetosphere.  We must consider the spin-down effect
 carefully in such a case.  In the polar region, the state
 $(\Lt\Omega_F)_{\rm min} < (\Lt\Omega_F)_\omega$ is easily obtained
 except for a non-rotating black hole case.    
 Then, the ingoing flows can carry the energy into the black hole with
 less angular momentum; this means the spin-down of the black hole.
 When disk's gases fall into the black hole along the disk--black hole
 connected dipole-like magnetic field lines
 \cite{Nitta-TT91,TT01}, such a situation would be possible.  Thus, the
 configuration of the magnetic field would be very important in
 astrophysical situations.          
 To estimate the actual spin-up/down of the black hole, we need to
 integrate the spin-up/down effects on each magnetic flux tube from the
 pole to the equator.  To do this, it is necessary to construct a
 reasonable model of magnetized accretion disk for distributions 
 $E=E(\theta_H)$, $L=L(\theta_H)$, $\Omega_F=\Omega_F(\theta_H)$ and
 $\eta=\eta(\theta_H)$. Furthermore, the function
 $\chi=\chi(\theta_{H})$ over the event horizon should be obtained from
 the studies of the GS equation around the event horizon.  Then, we can
 estimate the evolution of the rotating black hole.  However, this
 problem remains as our future task.

\section{ Concluding Remarks }           \label{sec:conc}

 We have discussed about stationary and axisymmetric ideal MHD inflows
 onto a black hole.  
 We have shown that a {\it non-singular}\/ distribution of
 $\beta(r,\theta)$ gives ingoing and/or outgoing trans-magnetosonic flow
 solutions without critical behavior.  Such solutions automatically
 satisfy the regularity conditions at the magnetosonic point, and passes
 through the magnetosonic point.  In a cold flow case, we only solve the
 quadratic equation for $M^2$ to obtain trans-fast magnetosonic flow
 solutions.  The ingoing flow passes through the inner or middle fast
 magnetosonic point. 
 When we consider a hot MHD flow, we can also treat the flow solution as
 a polynomial of high degree in $M^2$ without the regularity condition
 at the fast and slow magnetosonic points.

 In this paper, to discuss cold MHD inflows onto a black hole, we have
 not specified the function $\beta$ (or $\xi$) 
 without the GS equation. However, we have found the restrictions on 
 the field-aligned flow parameters under a given field geometry.  
 The ranges of possible $\Lt\Omega_F$ values for black hole accretion 
 are restricted by the condition at the Alfv\'en point as discussed in
 \S~\ref{sec:alfven} and Appendix~\ref{sec:A-star} and by the condition
 $A(r;\Psi)>0$ along the magnetic stream function that is discussed in
 \S~\ref{sec:super-fast}, where the function $A$ depends on the MHD flow
 energy $E$, the field geometry $\xi$ (or $\beta$), and the angular
 momentum related parameter  $\Lt\Omega_F$.   
 Although the value of $\Lt\Omega_F$ is related to the energy and
 angular momentum of the ingoing flow, the field geometry around the
 event horizon restricts these values.  
 Furthermore, we have discussed the inclination $\chi$ of magnetic field
 lines at the event horizon.  When we try to solve the magnetic field 
 distribution from the plasma source to the event horizon in a black
 hole magnetosphere, we should be aware of this boundary condition.  
 Note that the restrictions and conditions discussed in this paper
 are model independent as we are focusing on stationary and axisymmetric 
 black hole ideal MHD accretion. 
 In future studies of black hole magnetospheres (including numerical
 studies), our approach will be helpful to check the magnetic field
 structure near the event horizon.

 With the restriction on $L/E$ for MHD accretion, we have discussed the
 secular evolution of black hole's rotation for certain magnetic
 flux-tubes.   
 Although we have applied the boundary condition at the event horizon,
 our approach by the $\xi$-model guarantees that the ingoing MHD flow
 onto a black hole is trans-fast magnetosonic.  Then, we find that there
 are two asymptotic states for the spin evolution.  One is the state
 that (i) the angular velocity of the black hole approaches to that of
 the magnetic field line, $\omega_{H} \to \Omega_F$, and the other is
 the state that (ii) the spin-up due to angular momentum influx and the
 spin-down due to ingoing mass influx of MHD flows are canceled, where 
 $\tilde{L}\Omega_F = (\tilde{L}\Omega_F)_\omega$ is achieved.   
 For type~I case the asymptotic state of $\delta\omega_{H} = 0$ is
 achieved when $\tilde{L}\Omega_F = (\tilde{L}\Omega_F)_\omega$ (see
 Fig.~\ref{fig:spin}a), although the state (i) of $\omega_{H} \to
 \Omega_F$ may be realized before the state (ii). In the case of type~II 
 (see Fig.~\ref{fig:spin}b), when $L>0$ the asymptotic state is given by
 $\tilde{L}\Omega_F = (\tilde{L}\Omega_F)_\omega$, while the spin
 terminates as $\omega_{H} \to \Omega_F$ when $L<0$.

 When we discuss the actual asymptotic state in astrophysical
 situations, we should integrate the magnetic flux-tubes over the event
 horizon.  Although we can speculate on the final stage that the angular
 velocity of the black hole will be settled to a typical value of that
 of the whole magnetic field lines connected to the event horizon, we
 need to construct a reasonable MHD accretion model under a certain
 magnetic field configuration to specify the functions $L(\Psi)$ and
 $E(\Psi)$.  The basic idea of the black hole--disk connection of
 magnetic field lines has been suggested by some authors (e.g.,
 \cite{Hirotani92,Gammie99,GA97,Li02,Li04,PL03}).     
 The realistic features of magnetic field lines have been presented by    
 \cite{Nitta-TT91,TT01,Beskin97,Ghosh00,Uzdensky04,Uzdensky05}.  
 In an astrophysical point of view, the magnetic connection between the
 black hole and the disk would play a very important role.  The power
 input/output from the black hole depends on the ingoing MHD flow
 properties and the shape of the magnetic field lines.  The released
 energy of the plasma in the deep gravitational potential well can be 
 carried directly from the distant disk surface to the horizon by the
 black hole--disk connecting magnetic field lines.  Furthermore, when  
 the black hole is rapidly rotating, the rotational energy of the black 
 hole can be carried from the black hole to the disk through magnetic 
 interactions.  Such energy and angular momentum transport from the
 disk to the black hole determines the fate of hole's rotation.

 Although in this paper a stationary magnetosphere is discussed, their 
 dynamical phase is also an important problem.  The structure of a
 magnetosphere around the black hole is also discussed in
 \cite{KMSK00,Hirose04,MG04,McKinney06} by general relativistic MHD
 numerical simulations.  
 Our current investigations performed by analytical methods can help
 gain deeper insight from the results obtained by time-dependent general 
 relativistic MHD simulations.

%

\begin{acknowledgments} 
 M.T. would like to thank Sachiko Tsuruta and Rohta Takahashi for their 
 helpful comments.  
 This work was supported in part by the Grants-in-Aid of the Ministry 
 of Education, Culture, Sports, Science and Technology of Japan
 (17030006, 19540282, M.T.).  
\end{acknowledgments}


\appendix

\section{ Bending Angle of Magnetic Field Lines }     \label{sec:beta}

 We will discuss the conditions on the electric-to-magnetic field
 amplitude function $\xi$ or the bending angle $\beta$ along a magnetic
 field line connected between the disk surface and the black hole.  For
 a given function $\xi^2=\xi^2(r;\Psi)$ or $\beta^2=\beta^2(r;\Psi)$ 
 we can easily obtain a trans-fast magnetosonic accretion solution
 $M^2=M^2(r;\Psi)$ by the quadratic equation as discussed in
 \S~\ref{sec:trans-f}.  To obtain a physically acceptable accretion
 solution,  however, we should discuss some requirements on the function
 $\xi$ at some characteristic radii, where some conditions may restrict
 the functional form of $\xi^2$.    
 To discuss these requirements on $\xi^2$, we will consider a disk --
 black hole connecting magnetic field line, where the plasma injected
 from the disk surface streams toward the black hole along a magnetic
 field line.  First, the magnetic field line would almost corotate with
 the ``footpoint'' on the plasma source, where the magnetic field line
 is anchored.  So we may expect $B_\phi\sim 0$ ($\beta^2\gg 1$) there
 when the toroidal surface current dominates the poloidal one, or
 $\beta^2\sim O(1)$ when the toroidal surface current is almost the 
 same as the poloidal one.

 In the cases of type~I and ~III, the ingoing plasma injected from the 
 disk surface may bend the magnetic field toward the counterrotating
 direction (the trailed-shape of the magnetic field line: $B_\phi<0$)
 due to the plasma inertia effect. This situation is analogous to pulsar
 magnetosphere, although the magnetic field lines extend inwardly to
 the black hole. Conversely, the plasma flow ejected toward the rotation 
 direction (the leading-shaped: $B_\phi>0$) may be also possible.  In
 this case, however, the magnetic field line flips over at the ``anchor 
 point'' (see also \cite{Punsly01}), which corresponds to the Alfv\'en
 radius of $M^2\neq \alpha$ (where $B_\phi=0$). 
 At the anchor point, we see that $\xi^2=\infty$ and $\beta^2=0$.  
 At the Alfv\'en point that is the Alfv\'en radius with the condition 
 $M^2=\alpha$, we see $B_\phi\neq 0$ and both $\xi^2$ and $\beta^2$ have 
 finite values.

 Next, in type~II case, the rotating black hole must bend the magnetic
 field toward the rotating direction ($B_\phi>0$). In this type, no
 anchor point exists, but the ``corotation point'' where
 $\Omega_F=\omega$ exists between the inner and outer light surface,
 while in type~I and~III cases no corotation point exists on the way to
 the event horizon.  The value of $\xi$ at the corotation point, where
 $Y=0$, must become zero (i.e., $1+X=0$) to avoid the $A<0$ region near
 the corotation point, while $\beta^2$ has a finite value there. 
 [If not so, we obtain $(\Lt\Omega_F)^{+}= (\Lt\Omega_F)^{-}=0$ at the 
 corotation point, and we see the $A<0$ region between the plasma source
 and the event horizon; that is, no physical ingoing solution exists.]    
 Finally, at the event horizon, the boundary condition there requires
 that $\xi^2_{H} = 1$ or $\beta^2_{H} = -\alpha_{H}$. 
 These requirements on $\xi$ at characteristic radii are summarized in
 Table~\ref{tab:xi}.

\begin{table}[t]
\begin{center}
\caption{ Restrictions on the functions $\xi^2$ and $\beta^2$ at various
          characteristic radii are summarized.  
          The radius appeared in accretion solution is marked by
          ``$\circ$'' for types~I/III and~II, while the mark ``$\times$''
          shows the absence of that radius in the solution.
          \label{tab:xi}
}
\medskip
\begin{tabular}{llcccc} 
 \tableline\tableline
 \multicolumn{2}{c}{ Characteristic  Radii ~~} 
 & ~~~~$\xi^2$~~~~ & ~~~~$\beta^2$~~~~ & ~I/III~ & ~II~ \\ 
 \tableline
 Event Horizon (H) & $r_{H}$~~~~~ 
 & $1$ & $-\alpha_{H}$  & $\circ$ & $\circ$ \\
 Corotation point (C) & $ r_{\omega}$ 
 & $ 0 $ & finite   &  $\times$ & $\circ$  \\ 
 Alfv\'en Point (A) & $r_{\rm A}$ 
 & finite & finite  & $\circ$ & $\circ$  \\
 Anchor point (AN) & $r_{\rm AN}$ 
 & $\infty$ &  0   &  $\circ$ & $\times$ \\
 Plasma source (I) & $r_{I}$ 
 &  finite  &  finite  & $\circ$ & $\circ$  \\
 \tableline
\end{tabular}
\end{center}
\end{table}

 For example, for type~I and~III cases, we can simply set a model with
 $\xi^2(r,\theta) = \xi_{H}^2 = 1$ throughout the flow region considered
 for ingoing flows without the anchor point.  (Here, we consider a
 situation that the plasma source is placed inside the anchor point.) 
 As another example, we may add a deviation term from the
 $\xi^2(r,\theta)=1$ model, which must become zero on the event horizon, 
 as follows;    
 \begin{equation}
       \xi^2 = 1 + C_{\rm I} \left( \frac{\Delta}{\Sigma} \right) \ , 
	   \label{eq:xi_Iin}
 \end{equation}
 where $C_{\rm I}$ is a constant.

 To see the behaviors of accretion onto the black hole in type~II case,
 however, we should consider the function $\xi$ to satisfy both the
 boundary condition at the event horizon and the requirement on the
 corotation point.  Then, for example, we will consider the following
 function for $\xi$,   
 \begin{equation}
    \xi^2 = 
    \left[ 1 + C_{\rm II} \left( \frac{\Delta}{\Sigma} \right) \right] 
    \left( \frac{\omega-\Omega_F}{\omega_{H}-\Omega_F} \right)^2 \ ,  
	   \label{eq:xi_IIin}
 \end{equation}
 where $C_{\rm II}$ is a constant. 
 [The functions $(\Lt\Omega_F)^\pm$ seen in Figure~\ref{fig:A-region}d
 could be plotted under this distribution of $\xi$.]   
 Note that the $\xi^2(r,\theta)=1$ inflow model in the type II case
 requires $\beta=0$ at the corotation point. This requirement means
 $B_\phi=0$ that corresponds to the anchor point, but there is {\it
 no}\/ anchor point in type II flow solutions. Thus, the simple
 $\xi^2(r,\theta)=1$ model can not be applied as an inflow solution for
 a rapidly rotating black hole (type~II) case. 

 For outflows ($r\gg m$), for example, we can make a model 
 \begin{equation}
   \xi^2 =  1 - \frac{1}{\EE^2} + \zeta_0 \  ,      \label{eq:xi_out}
 \end{equation}
 where $\zeta_0$ is the parameter for the magnetic field geometry. 
 For $\zeta_0 < 0$, the outgoing flow reaches the distant region with a 
 finite Mach number, while for $\zeta_0 > 0$ the flow is confined within
 a certain radius $R_{c}$ where the Mach number becomes to diverge (see
 \cite{TT03}, for detail discussions).

 In a realistic situation for the black hole magnetosphere, a global
 structure of magnetic field lines would not be able to be expressed by
 simple forms of function $\xi$.  To understand the basic properties of
 MHD inflows/outflows, it will be helpful to use function
 (\ref{eq:xi_Iin}) or (\ref{eq:xi_IIin}) for ingoing winds and function 
 (\ref{eq:xi_out}) for outgoing winds.  
 In \S~\ref{sec:trans-f}, we show Figures~\ref{fig:acc}
 and~\ref{fig:acc-Kerr} that are accretion solutions {\it along a given
 stream line}\/ with a given function $\xi(r;\Psi)$.  However, we should
 note that, as a practical matter, we can obtain {\it the velocity
 distribution}\/ $u^r(r,\theta)$ in the poloidal plane when the function
 $\xi(r,\theta)$ is specified as a model. 

 Now, we consider the restriction on the function $\xi(r,\theta)$ for
 the $A>0$ region, which is the necessary condition for MHD
 ingoing/outgoing winds, in the black hole magnetosphere.  
 If the discriminant in equation (\ref{eq:c-jj}) becomes zero at some
 location ($r, \theta$) in the super-Alfv\'enic region,  equation
 (\ref{eq:c-jj}) becomes 
 \begin{equation}
    (\Lt\Omega_F)^{+} = (\Lt\Omega_F)^{-} = (\Lt\Omega_F)_{d} \ , 
                                                      \label{eq:Y_d=0} 
 \end{equation}
 where 
 \begin{equation}
    (\Lt\Omega_F)_{d} \equiv 1-\frac{G_t}{\EE^2} \ ,
                                                        \label{eq:Y_d}  
 \end{equation}
 and the corresponding $\xi^2$ value, $\xi^2_{d}$, is expressed
 as  
 \begin{equation}
     \xi^2_{d}(r,\theta) \equiv 
     \frac{g_{\phi\phi}(\Omega_F-\omega)^2(g_{tt}-\EE^2)} 
          {G_t^2 - \alpha \EE^2} \ .                     \label{eq:xi-A}
 \end{equation}
 Under the condition of $\xi^2(r; \Psi) < \xi_{d}^2(r; \Psi)$ along a
 magnetic field line $\Psi=\Psi(r,\theta)$, the requirment $A > 0$ on
 the super-Alfv\'{e}nic solution is possible for a suitable $\Lt\Omega_F$
 value at most in the magnetosphere.  
 If the situation with $\xi^2(r;\Psi) = \xi^2_{d}$ is achieved on the
 magnetic field line considered, the $A=0$ surface must appear on
 the way to the $\xi^2 = \xi^2_{d}$ radius (i.e., between the plasma
 source and the location with $\xi^2=\xi^2_{d}$).  Such a situation
 gives an unphysical flow solution.    
 Note that, for the inflow, if the $\xi^2=\xi^2_{d}$ surface encloses
 the event horizon, no black hole accretion solution exists.  
 Although the value of $\xi^2_{d}$ diverges ($\xi^2_{d}=\pm\infty$) at
 surfaces with $\alpha/G_t^2 = 1/ \EE^2$ [ or $Y = (\Lt\Omega_F)_{d}$],  
 the inner and outer surfaces of $\xi^2_{d}(r,\theta)=\infty$ exist in
 the black hole magnetosphere.   
 Between these surfaces, the value of $\xi_{d}^2$ is always negative 
 and the discriminant in equation (\ref{eq:c-jj}) is positive; that is,
 the condition $A>0$ is achieved everywhere.   
 This $\xi_{d}^2<0$ region occupies the magnetosphere at most between
 the inner and outer light surfaces.  For the $\EE\gg 1$ flow, the
 $\xi^2_{d}(r,\theta)=\infty$ surfaces coincide with the light surfaces
 (given by $\alpha = G_t^2/\EE^2 \sim 0$).   
 Thus, for accretion/wind solutions, the condition $\xi^2(r; \Psi) <
 \xi_{d}^2(r; \Psi)$ must be required along a magnetic field line
 considered, except for the $\xi_{d}^2<0$ region.

\section{The restriction on the Alfven point by the $A>0$ condition } 
\label{sec:A-star}

 Here, we consider the restriction on the Alfv\'en point due to the  
 condition $A>0$ for the MHD accretion solution again (see
 \S~\ref{sec:alfven}).  Although the relation $\Lt\Omega_F = Y_{\rm A}$
 gives the Alfv\'{e}n surface, the Alfv\'en surface must be located
 within the $A>0$ region. Then, the innermost limit of the inner
 Alfv\'{e}n point (marked by A$\ast$) for a given magnetic flux surface
 is given by $Y_{{\rm A}\ast} = (\Lt\Omega_F)^{+}_{{\rm A}\ast}$ for
 type~I and~III, and $Y_{{\rm A}\ast} = (\Lt\Omega_F)^{-}_{{\rm A}\ast}$
 for type II, whose relation can be reduced to  
 \begin{equation}
     (\Lt\Omega_F)_{\rm A \ast} \equiv 
     Y_{{\rm A}\ast} = 1-\frac{(G_t)_{\rm A \ast}}{\EE^2} \ . 
     \label{eq:mia}
 \end{equation}
 Note that this relation is independent of the function $\xi$. 
 If the Alfv\'en point is located inside this critical location, the
 location with $A=0$ appearers in the super-Alfv\'enic region. Then,   
 the requirement for the acceptable $\Lt\Omega_F$ range discussed in 
 \S~\ref{sec:alfven} should be modified to avoid such a situation (see,
 Figs~\ref{fig:A-region}c and~\ref{fig:A-region}d, where the location of
 $r=r_{\rm A \ast}$ is given by the fill circle on the $\Lt\Omega_F = Y$ 
 curve).  That is, we should take the range with 
 $(\Lt\Omega_F)_{\rm A \ast} < \Lt\Omega_F$ for the $E<0$ inflows 
 in the type~II case, where $(\Lt\Omega_F)_{\rm A \ast} > 1$.  
 On the other hand, we should take the range with 
 $\Lt\Omega_F < (\Lt\Omega_F)_{\rm A \ast}$ for the inflows ($E>0$) in
 the type~I and~III cases,  where $(\Lt\Omega_F)_{\rm A \ast} < 1$.

 Thus, with respect to $\Lt\Omega_F$ for the black hole accreting flows,
 we summarize the necessary condition with        
 \begin{equation}
     | 1 - \Lt\Omega_F | > \frac{ |({G_t})_{\rm A \ast}| }{\EE^2} \ . 
                                                        \label{eq:Ya-1}
 \end{equation}
 If not so, $A=0$ is realized between the inner Alfv\'{e}n point and the
 event horizon; that is, the Mach number of the trans-Alfv\'{e}nic
 solution diverges there (unphysical).  When we consider the
 trans-Alfv\'{e}nic ingoing flow, for $r_{\rm A} > r_{\rm A \ast}$
 equation (\ref{eq:Ya-1}) can be rewritten as $\ee^2 > \alpha_{\rm A}$,
 which gives a physical solution.  However, no physical
 trans-Alfv\'{e}nic flow, where $u_p^2(r_{\rm A})<0$, is obtained for
 $r_{\rm A} \leq r_{\rm A \ast}$.

 When we treat a magnetically-dominated flow with 
 $\EE\gg 1$, we should mention the deviation from the limit of 
 large magnetization; i.e., the force-free case.  (In the 
 magnetically-dominated flow limit, we see that $\EE\to\infty$, 
 $\Lt\Omega_F \to 1$ and $r_{\rm A \ast} \to r_{L}^{\rm in}$.)    
 For the magnetically-dominated flow, the deviation from the force-free
 model with respect to the $\Lt\Omega_F$ value is of the order of
 $\EE^{-2}$, and the location of the Alfv\'{e}n point also separates
 from the inner light surface in the same order 
 [~$ (r_{\rm A \ast}/r_L) = 1 + |O(\EE^{-2})| $~].     
 When we discuss some problems (e.g., the energy and angular momentum 
 transport) in the MHD framework, the deviation from the force-free
 model becomes important.  
 This necessary condition for the trans-Alfv\'enic flow solutions
 discussed above also applies to the outflows.

\section{ The nature of the parameter $\chi$ } \label{sec:chi}

 In \ref{sec:inc}, we introduced the function $\chi$ that represents
 the magnetic field configuration near the event horizon.  Although the  
 function $\chi(\theta_{H})$ should be given by solving the GS equation,
 in this paper we only treat it as a parameter without the GS equation.   
 Here, we discuss the characteristic nature of the parameter $\chi$.

 First, we consider the maximum $\chi$ value ($\equiv \chi_{\rm max}$)
 for existing $A>0$ region in both type~I and~II cases, where we obtain      
 \begin{equation}
   \chi_{\rm max} \equiv
   \frac{1+[(\Omega_F^2-\omega_{H}^2)/\omega_{H}^2](g_{tt}^{H}/\EE^2)}
   {{\cal H}[(\Omega_F-\omega_{H})/\omega_{H}](1-g_{tt}^{H}/\EE^2)} \ , 
 \end{equation}
 which gives $(\Lt\Omega_F)_{H}^{+}=(\Lt\Omega_F)_{H}^{-}$.  
 In the Schwarzschild black hole case, we have $\chi_{\rm max} \sim
 (2m\Omega_F\sin\theta)^{-2}$, which is finite near the equator. 
 When  $\chi > \chi_{\rm max}$, no physical trans-fast magnetosonic
 accretion solution exists for any $\Lt\Omega_F$ values.  On the other
 hand, we have that $(\Lt\Omega_F)_{H}^{+} \sim (\Lt\Omega_F)_{H}^{-}
 \sim \Omega_F/\omega_{H} = Y_{H}$ when $(-\chi) \gg 1$.

 Second, we see that the value of $(\Lt\Omega_F)_{H}^{+}$ diverges when 
\begin{equation} 
  \chi = \chi_{\infty} 
  \equiv -(\omega_{H}+\Omega_F)/({\cal H}\omega_{H}) \ ,  
\end{equation}
 while $(\Lt\Omega_F)_{H}^{-}$ is finite.  For a rapidly rotating black
 hole case (type~II) this situation, $\chi=\chi_\infty$, arises in the
 $\chi>0$ region (see Fig.~\ref{fig:Y_H}b).  For a slowly rotating black
 hole case (type~I), however, the value of $\chi_\infty$ is always
 negative, and then no restriction by the $A_{H}<0$ condition exists on
 the acceptable $\Lt\Omega_F$ range (see Fig.~\ref{fig:Y_H}a).  
 Furthermore, we also specify the third value  
\begin{equation}
 \chi_{0} \equiv - [(r_{H}-m)\Sigma_{H}/(2 m^2 r_{H})]/ \EE^2 \ ,  
\end{equation}
 where the function $(\Lt\Omega_F)_{H}^{+}$ becomes zero.   
 Although the function $(\Lt\Omega_F)_{H}^\pm$ has a $\EE$ dependence,
 the dependence is weak for $\EE > 1$.  In the limit of $\EE=\infty$,
 we see that $\chi_{0} = 0$, where $(\Lt\Omega_F)_{H}^{+} = 0$, and
 $\chi_{\rm max} = \chi_\infty \omega_{H}^2/(\Omega_F^2-\omega_{H}^2)$, 
 where $(\Lt\Omega_F)_{H}^\pm = 1$, while $\chi_\infty$ is independent
 of $\EE$.

 In \S \ref{sec:super-fast} we introduce the restrictions on
 $\Lt\Omega_F$ by the condition (\ref{eq:Ya-1}) at the Alfv\'en point.  
 The additional restriction from the boundary condition at the event
 horizon should be also imposed to avoid the $A=0$ surface on the
 super-Alfv\'enic accretion flow.   
 Then, the ingoing flow solution can be classified by the $\chi$ value. 
 First, for the slowly rotating black hole case (type I), the ranges of   
 $\Lt\Omega_F < (\Lt\Omega_F)_{H}^{-}$ and 
 $\Lt\Omega_F > (\Lt\Omega_F)_{H}^{+}$ are 
 forbidden from the boundary condition at the event horizon.  In addition
 to this restriction, the range of $\Lt\Omega_F < (\Lt\Omega_F)_{\rm min}$ 
 and $(\Lt\Omega_F)_{\rm A \ast} < \Lt\Omega_F$ are also forbidden from the
 condition of the Alfv\'{e}n point, where $(\Lt\Omega_F)_{\rm A \ast} < 1 $
 for type~I case.  When $ \chi < \chi_{\rm max}$, the acceptable range of
 $\Lt\Omega_F$ is given by  
 $\mbox{MAX}[(\Lt\Omega_F)_{\rm min}, (\Lt\Omega_F)^{-}_{H} ] <
 \Lt\Omega_F < (\Lt\Omega_F)_{\rm A \ast} $ for the magneto-like accretion
 solution, while that is $\mbox{MAX}[ (\Lt\Omega_F)_{\rm min},
 (\Lt\Omega_F)^{-}_{H} ] < \Lt\Omega_F < (\Lt\Omega_F)^{+}_{\rm min} <
 (\Lt\Omega_F)_{\rm A \ast} $  for the hydro-like accretion solution, where
 $(\Lt\Omega_F)^{+}_{\rm min}$ is the minimum of the function
 $(\Lt\Omega_F)^{+}=(\Lt\Omega_F)^{+}(r)$ (see also
 Fig.~\ref{fig:A-region}c).  Note that $(\Lt\Omega_F)^{-}_{H}<0$ for $
 \chi < \chi_{0}$, while the value of $(\Lt\Omega_F)^{\rm min}$ is
 always positive.   For type III case, we can see the similar behavior 
 of $\Lt\Omega_F$.

 Next, for a rapidly rotating black hole case (type II), the value of
 $\chi$ is related to the signature of the total energy $E$ and angular
 momentum $L$ of the flow.  When $\chi < \chi_\infty$ the accretion
 solution in the range $(\Lt\Omega_F)_{H}^{+} < \Lt\Omega_F <
 (\Lt\Omega_F)_{H}^{-}$  must be forbidden from the boundary    
 condition at the event horizon, and when  
 $\chi_\infty < \chi < \chi_{\rm max}$ the solution in the ranges
 $\Lt\Omega_F < (\Lt\Omega_F)_{H}^{-}$ and 
 $\Lt\Omega_F > (\Lt\Omega_F)_{H}^{+}$ must be forbidden.  
 Furthermore, the range  $(\Lt\Omega_F)_{H}^{-} < \Lt\Omega_F <
 (\Lt\Omega_F)_{\rm A \ast}$, where $(\Lt\Omega_F)_{\rm A \ast} > 1$
 for accretion solutions in type~II, is also forbidden by the existence
 of the $A=0$ surface between the Alfv\'en point and the event horizon as
 mentioned in \S \ref{sec:super-fast}.  Then, we obtain the acceptable
 ranges as $\Lt\Omega_F < (\Lt\Omega_F)_{H}^{+}$ and 
 $\Lt\Omega_F > (\Lt\Omega_F)_{\rm A \ast}$ for ingoing flows with 
 $\chi < \chi_\infty$, while the acceptable range is 
 $(\Lt\Omega_F)_{\rm A \ast} < \Lt\Omega_F < (\Lt\Omega_F)_{H}^{+}$ for 
 inflows with $\chi_\infty < \chi < \chi_{\rm max}$.

 Here, we will consider three ranges of $\chi$-value for type II case; 
 that is, 
\begin{description}
 \item[(a)]  $\chi_\infty < \chi < \chi_{\rm max}$, 
 \item[(b)]  $\chi_{0} < \chi < \chi_\infty$, 
 \item[(c)]  $ \chi < \chi_{0}$.  
\end{description}
 In all cases, the negative energy inflow is acceptable. In the cases
 (b) and (c) the positive energy inflow is also possible, while in
 the case (a) energy $E$ must be negative.  Thus, the value
 $\chi_\infty$ gives the maximum $\chi$ value for the positive energy
 input to the black hole. 
 Similarly, for the cases (a) and (b), the angular momentum of the
 inflows must be negative, while in the case (c) both the positive and 
 negative inflows are available.  The $\chi=0$ inflow is included in the
 case (b); the example of $(\Lt\Omega_F)=(\Lt\Omega_F)^\pm(r)$ curve
 shown in Figure~\ref{fig:A-region}b corresponds to this case.     
 When we consider the $a\to m$ limit, where ${\cal H}\to \infty$, we
 obtain $(\Lt\Omega_F)_{H}^\pm(\chi) \to Y_{H} = 2m\Omega_F$, and  
 $\chi_{\rm max} \to 0$,  $\chi_\infty \to 0$ and  $\chi_0 \to 0$. Then, 
 only $\chi < 0$ is acceptable, which gives the acceptable $\Lt\Omega_F$
 of $\Lt\Omega_F < Y_{H}$ and $(\Lt\Omega_F)_{\rm A \ast} < \Lt\Omega_F$.   
 When $\omega_{H} \to \Omega_F$, we have ${\cal H} = 0$, 
 $\chi_{\rm max} \to \infty$,  $\chi_\infty \to \infty$,  
 $(\Lt\Omega_F)_{H}^{-}(\chi)=1$ and $(\Lt\Omega_F)_{H}^{+}(\chi) = 0$.
 The acceptable $\Lt\Omega_F$ ranges are $\Lt\Omega_F < 0$ and
 $(\Lt\Omega_F)_{\rm A \ast} < \Lt\Omega_F$; only negative angular
 momentum flows are available.         
 For the off-equatorial inflow of $\theta\ll 1$, we have 
 ${\cal H}\propto \sin^2\theta$, $|\chi_\infty|\propto \sin^{-2}\theta$
 and $\chi_{\rm max}\propto \sin^{-2}\theta$, while the value of
 $\chi_\infty$ does not change drastically.  Just on the pole, we find 
 that $(\Lt\Omega_F)_{H}^{+}\sim 0$ and that for the type~II case only
 the inflow with $L\leq 0$ is available. 
 Thus, the ratio of the toroidal and poloidal magnetic fields that is
 related to the bending angle of the magnetic field line is restricted  
 by the total energy and angular momentum of accretion onto a rotating
 black hole.

%


\begin{thebibliography}{} 
\bibitem[Balbus \& Hawly(1998)]{BB98} 
         S.~A.~Balbus, \& J.~F.~Hawly,  Rev. Mod. Phys., 70, 1 (1998)
\bibitem[Meier(2005)]{Meier05}
         D.~L.~Meier,  astro-ph/0504511 (2005)
\bibitem[Ghosh \& Lamb(1979)]{GL79}
         P.~Ghosh, \& F.~K.~Lamb,  ApJ, 234, 296 (1979)
\bibitem[Bardeen(1970)]{Bardeen70} 
         J.~M.~Bardeen,  Nature, 226, 64 (1970)
\bibitem[Thorne(1974)]{Thorne74}
         K.~S.~Thorne,  ApJ, 191, 507 (1974)
\bibitem[Blandford \& Znajek(1977)]{BZ77} 
         R.~D.~Blandford, \& R.~L.~Znajek,  MNRAS, 179, 433 (1977)
\bibitem[Takahashi et al.(1990)]{TNTT90}
         M.~Takahashi, S.~Nitta, Y.~Tatematsu, \& A.~Tomimatsu, 
         ApJ, 363, 206 (1990) 
\bibitem[Park \& Vishniac(1988)]{PV88}
         S.~J.~Park, \& E.~T.~Vishniac,  ApJ, 332, 135 (1988)
\bibitem[Weber \& Davis(1967)]{WD67}
         E.~J.~Weber, \& L.~Davis, Jr.,  ApJ, 148, 217 (1967)
\bibitem[Camenzind(1986)]{Camenzind86} 
         M.~Camenzind,  A \& A , 162, 32 (1986)
\bibitem[Takahashi(2002)]{Takahashi02} 
         M.~Takahashi,  ApJ, 570, 264 (2002)
\bibitem[Nitta, Takahashi, \& Tomimatsu(1991)]{Nitta-TT91}
         S.~Nitta, M.~Takahashi, \& A.~Tomimatsu,  Phys. Rev., D44,
         2295 (1991)  
\bibitem[Tomimatsu \& Takahashi(2003)]{TT03} 
         A.~Tomimatsu, \& M.~Takahashi,  ApJ, 592, 321 (2003)
\bibitem[Bardeen, Press, \& Teukolsky(1972)]{BPT72} 
         J.~M.~Bardeen, W.~H.~Press, \& S.~A.~Teukolsky, 
         ApJ, 178, 347 (1973)  
\bibitem[Tomimatsu \& Takahashi(2001)]{TT01} 
         A.~Tomimatsu, \& M.~Takahashi, ApJ, 552, 710 (2001)
\bibitem[Hirotani et al.(1992)]{Hirotani92}
         K.~Hirotani, M.~Takahashi,  S.~Nitta, \& A.~Tomimatsu, 
         ApJ, 386, 455 (1992)
\bibitem[Gammie(1999)]{Gammie99}
         C.~F.~Gammie,  ApJ, 522, L57 (1999)
\bibitem[Ghosh \& Abramowicz(1997)]{GA97}
         P.~Ghosh, \& M.~A.~Abramowicz,  MNRAS, 292, 887 (1997)
\bibitem[Li(2002)]{Li02}
         L.~-X.~Li,  ApJ, 567, 463 (2002)
\bibitem[Li(2004)]{Li04}
         L.~-X.~Li,  PASJ, 56, 685 (2002)
\bibitem[von Putten \& Levinson(2003)]{PL03}
         M.~H.~P.~M.~van Putten and A.~Levinson,  ApJ, 584, 937 (2003)
\bibitem[Beskin(1997)]{Beskin97} 
         V.~S.~Beskin, Physics-Uspekhi, 40, 659 (1997)
\bibitem[Ghosh(2000)]{Ghosh00}
         P.~Ghosh,  MNRAS, 315, 89 (2000)
\bibitem[Uzdensky(2004)]{Uzdensky04}
         D.~A.~Uzdensky,  ApJ, 603, 652 (2004)
\bibitem[Uzdensky(2005)]{Uzdensky05}
         D.~A.~Uzdensky,  ApJ, 620, 889 (2005)
\bibitem[Koide et al.(2000)]{KMSK00}
         S.~Koide, D.~L.~Meier, K.~Shibata, \& T.~Kudoh, 
         ApJ, 536, 668 (2000)
\bibitem[Hirose et al.(2004)]{Hirose04}
         S.~Hirose, J.~H.~Krolik, J.~P.~De Villiers, \& J.~H.~Hawley, 
         ApJ, 606, 1083  (2004)
\bibitem[McKinney \& Gammie(2004)]{MG04}
         J.~C.~McKinney, \& C.~F.~Gammie,  ApJ, 611, 977 (2004)
\bibitem[McKinney(2006)]{McKinney06}
         J.~C.~McKinney,  MNRAS, 368, 1561 (2006)
\bibitem[Punsly(2001)]{Punsly01}
         B.~Punsly, \textit{Black Hole Gravitohydromagnetics},
         (Springer, Berlin, 2001)
\end{thebibliography}
\end{document}